\documentclass{aa}  
\usepackage{graphicx}
\usepackage{txfonts}
\usepackage{rotating}
\usepackage{multirow}

\newcommand{\avgg}[1]{\left< #1 \right>} 
\newcommand{\avg}[1]{#1} 
\def\obj{QSO~2237$+$0305}

\begin{document}

\title{Zooming into the broad line region of the gravitationally lensed quasar 
\vspace*{1mm}  \obj\ $\equiv$ the Einstein Cross
 \thanks{Based on observations made 
 with the ESO-VLT Unit Telescope \#~2 Kueyen 
 (Cerro Paranal, Chile; Proposals 
 073.B-0243(A\&B),
 074.B-0270(A), 
 075.B-0350(A), 
 076.B-0197(A),
 177.B-0615(A\&B), PI: F. Courbin).}
 }
\subtitle{III. Determination of the size and structure of the \ion{C}{IV} and \ion{C}{III]} emitting regions using microlensing}


\author{ D.\,Sluse\inst{1} \and R.\,Schmidt\inst{1} \and F.\,Courbin\inst{2}
  \and D.\,Hutsem\'ekers\inst{3} \and G.\,Meylan\inst{2},
  A.\,Eigenbrod\inst{2} \and T.\,Anguita \inst {4,5} \and E.\,Agol \inst{6} \and J.\,Wambsganss \inst{1}}


\institute{ Astronomisches Rechen-Institut am Zentrum f\"ur Astronomie
  der Universit\"at Heidelberg M\"onchhofstrasse 12-14, 69120
  Heidelberg, Germany \and Laboratoire d'Astrophysique, Ecole
  Polytechnique F\'ed\'erale de Lausanne (EPFL), Observatoire de
  Sauverny, 1290 Versoix, Switzerland \and F.R.S.-FNRS, Institut d'Astrophysique et
  de G\'eophysique, Universit\'e de Li\`ege, All\'ee du 6 Ao\^ut 17,
  B5c, 4000 Li\`ege, Belgium \and Centro de Astro-Ingenier\'ia,
  Departamento de Astronom\'ia y Astrof\'isica, P. Universidad
  Cat\'olica de Chile, Casilla 306, Santiago, Chile \and
  Max-Planck-Institut f\"ur Astronomie, K\"onigstuhl 17, 69117
  Heidelberg, Germany \and Astronomy Department, University of
  Washington, Box 351580, Seattle, WA 98195, USA }

\date{Received 9 November 2010 ; Accepted 13 December 2010}


\abstract
{}
   {We aim to use microlensing taking place in the lensed quasar \obj\, to study the structure of the broad line region and measure the size of the region emitting the \ion{C}{IV} and \ion{C}{III]} lines.}
   {Based on 39 spectrophotometric monitoring data points obtained between Oct. 2004 and Dec. 2007, we derived lightcurves for the \ion{C}{IV} and \ion{C}{III]} emission lines. We used three different techniques to analyse the microlensing signal. Different components of the lines (narrow, broad, and very broad) were identified and studied. We built a library of the simulated microlensing lightcurves that reproduce the signal observed in the continuum and in the lines provided only the source size is changed. A Bayesian analysis scheme is then developed to derive the size of the various components of the BLR.}
   {1. The half-light radius of the region emitting the \ion{C}{IV} line is found to be $R_{\ion{C}{IV}} \sim 66^{+110}_{-46}$ light-days = 0.06$^{+0.09}_{-0.04}$ pc = 1.7$^{+2.8}_{-1.1}$\,10$^{17}$ cm (at 68.3\% CI). Similar values are obtained for \ion{C}{III]}. Relative sizes of the carbon-line and V-band continuum emitting-regions are also derived with median values of $R^{\rm line}/R^{\rm cont}$ in the range 4 to 29, depending on the FWHM of the line component. \\  
   2. The size of the \ion{C}{IV} emitting region agrees with the radius-luminosity relationship derived from reverberation mapping. Using the virial theorem, we derive the mass of the black hole in \obj\, to be $M_{BH} \sim 10^{8.3 \pm 0.3} M_{\sun}$. \\
   3. We find that the \ion{C}{IV} and \ion{C}{III]} lines are produced in at least 2 spatially distinct regions, the most compact one giving rise to the broadest component of the line. The broad and narrow line profiles are slightly different for \ion{C}{IV} and \ion{C}{III]}. \\ 
   4. Our analysis suggests a different structure for the \ion{C}{IV} and \ion{Fe}{II+III} emitting regions, with the latter produced in the inner part of the BLR or in a less extended emitting region than \ion{C}{IV}. 
}
   {}

\titlerunning{Size and structure of the BLR in the Einstein Cross}
\authorrunning{D. Sluse et al.}


\maketitle

\section{Introduction}

We know that quasars and active galactic nucleii (AGN) are powered by
matter accreted onto a supermassive black hole. The accretion of
material in the direct vicinity of the central black hole releases
most of the quasar energy in the form of power-law continuum emission.
Ionised gas surrounds the central accretion disc and gives rise to
broad emission lines, which are used as footprints that allow the
identification and classification of quasars. Our knowledge of the
kinematics and physical conditions prevailing in the BLR gas remain
elusive, especially because the nuclear region of quasars is still
spatially unresolved with existing instrumentation.

Current insights into the BLR come from various kind of studies:
empirical modelling of the line shape with kinematical models, use of
photo-ionisation codes to reproduce the observed flux ratios between
spectral lines, spectropolarimetric observations, statistical study of
the width and asymmetry of the lines, use of the principal component
analysis technique, and velocity resolved reverberation mapping
(e.g. Boroson \& Green~\cite{BOR92}, Sulentic et al.~\cite{SUL00},
Smith et al.~\cite{SMI05}, Marziani et al.~\cite{MAR06}, Zamfir et
al.~\cite{ZAM08}, Gaskell~\cite{GAS09}, ~\cite{GAS10b}, Bentz et
al.~\cite{BEN10}). Despite the development and many successes of these
methods, as briefly summarised below, we still do not completely
understand the structure and kinematics of the BLR.  The microlensing
of the broad emission lines in multiply imaged lensed AGNs provides us
with a powerful alternative technique for looking at the BLR, measure its
size even in high luminosity distant quasars, and get hints of the
structure and geometry of both emission and intrinsic absorption
within the BLR (e.g. Schneider \& Wambsganss~\cite{SCH90},
Hutsem\'ekers et al.~\cite{HUT94}, Lewis \& Belle~\cite{LEW98}, Abajas
et al.~\cite{ABA02}, Popovi{\'c} et al.~\cite{POP03}, Lewis \&
Ibata~\cite{LEW04}, \cite{LEW06}, Richards et al.~\cite{RIC04},
Abajas et al.~\cite{ABA07}, Sluse et al.~\cite{SLU07},~\cite{SLU08},
Hutsem\'ekers et al.~\cite{HUT10}).

\subsection{Phenomenology of the line profiles}

Our primary clue on the BLR comes from the shape of the broad emission
lines. The most detailed studies of broad emission lines have focused
on two lines: \ion{H}{$\beta$}\,$\lambda$4863 and
\ion{C}{IV}$\,\lambda$1549 (Sulentic et al.~\cite{SUL00} for a
review).  Of direct interest for the present work is
\ion{C}{IV}$\lambda$1549. The \ion{C}{IV} profile shows a broad
variety of shapes from strongly asymmetric to symmetric (e.g. Wills et
al.~\cite{WIL93}, Baskin \& Laor \cite{BAS05}), has an equivalent
width anti-correlated with its intensity (Francis et
al.~\cite{FRA92}), and shows greater variability in the wings than in
the core (Wilhite et al.~\cite{WIL06}). In addition, the \ion{C}{IV}
line is also found to be systematically blueshifted by several hundred
to a few thousand kilometres per second compared to the low ionisation
lines (Gaskell~\cite{GAS82}, Corbin~\cite{COR90}, Vanden Berk et
al.~\cite{VAN01}). The analysis of about 4000 SDSS quasars by Richards
et al. (\cite{RIC02}) suggests that this shift is caused by a lack of
flux in the red wing of the line profile and correlates with the
quasar orientation. Variability studies and emission-line
decomposition techniques (principal component analysis and analytical
fitting) indicate that the region emitting \ion{C}{IV} could be built
up with two components, a ``narrow'' emission core of FWHM $\sim$ 2000
km\,s$^{-1}$ emitted in an intermediate line region (ILR) possibly
corresponding to the inner part of the narrow line region and a very
broad component (VBC) with FWHM $\sim$7000 km\,s$^{-1}$ producing the
line wings (Wills et al.~\cite{WIL93}, Brotherton et
al.~\cite{BRO94a}, Sulentic~\cite{SUL00}, Wilhite et
al.~\cite{WIL06}). In radio-quiet objects, the VBC is observed to be
systematically blueshifted by thousands of km\,s$^{-1}$ with respect
to the narrow core (Brotherton et al.~\cite{BRO94a},
Corbin~\cite{COR95}) suggesting it is associated with outflowing
material. The ILR component disclosed in \ion{C}{IV} is probably
different from the one recently uncovered in \ion{H}{$\beta$} (Hu et
al.~\cite{HU08}), but its exact nature and the physical conditions in
this region are still being debated (Brotherton et al.~\cite{BRO94b},
Sulentic \& Marziani~\cite{SUL99}, Marziani et al.~\cite{MAR06}).

The properties of the \ion{C}{III]} emission line have received less
attention in the literature, probably mostly because the line is
blended with \ion{Al}{III}\,$\lambda$1857,
\ion{Si}{III]}\,$\lambda$1892 and, an
\ion{Fe}{II+III}\,$\lambda$1914 complex. In their study of the
\ion{C}{IV} and \ion{C}{III]} emission, Brotherton et
al. (\cite{BRO94a}) find that the 2 lines often have different
profiles and that, for some objects, a two-component decomposition
(i.e. narrow core+very broad wings) does not provide a good model of
\ion{C}{III]}, so a third component is required. They also find that the
VBC needed to reproduce \ion{C}{III]} has to be larger than the
corresponding component of \ion{C}{IV}. This supports the idea that
the region emitting \ion{C}{III]} is different from the one emitting
\ion{C}{IV} (Snedden \& Gaskell~\cite{SNE99}). 

Recently, Marziani et al. (\cite{MAR10}), based on the line profile
decomposition of a small sample of AGNs selected in the 4D Eigenvector
1 context (e.g. Boroson \& Green~\cite{BOR92}, Zamfir et
al.~\cite{ZAM08}), suggest that all the broad emission line profiles
are composed of three components of variable relative intensity and
centroid shift (from line to line in a given object and between
objects). They suggest a classical unshifted broad component
(FWHM=600-5000 km\,s$^{-1}$), a redshifted very broad component and a
blueshifted component mostly visible in the so-called Population A
objects (i.e. objects with FWHM $<$ 4000 km\,s$^{-1}$). Based on line
ratios, they also tentatively infer that these components arise from
different emitting regions.

The smoothness of the line profiles and the physical conditions
derived from photo-ionisation models allowed several authors to put
constraints on the ``structure'' of the BLR gas. Two popular models
remain. The first one considers that the BLR is a clumpy flow composed
of small gas clouds, and the second one assumes a smooth gas outflow
originating in an accretion disc (Elvis ~\cite{ELV00}, Laor
~\cite{LAO07} and ref. therein). Several geometries have been
considered for the BLR gas, the most popular ones being disc-like,
spherical and biconical models (e.g. Chen \& Halpern~\cite{CHE89},
Robinson \cite{ROB95}). Murray and Chiang (\cite{MUR97}) demonstrate
that a continuous wind of gas originating in an accretion disc can
successfully reproduce the profile and systematic blueshift of the
\ion{C}{IV} emission line. Other indications that a fraction of the
BLR material has a disc-like geometry comes from statistical studies
of the broad line profiles in samples of (mostly radio-loud) AGNs
(Vestergard et al. ~\cite{VES00}, McLure \& Dunlop~\cite{MCL02},
Jarvis \& McLure~\cite{JAR06}, Decarli et al.~\cite{DEC08}, Risaliti
et al.~\cite{RIS10}), and from spectro-polarisation observations of
Balmer lines in AGNs (Smith et al.~\cite{SMI05}). Some of these
studies also suggest there is a second, spherically symmetric
component with Keplerian motion.  The spectropolarimetric observations
of PG 1700+518 by Young et al. (\cite{YOU07}) support a disc+wind
model for the \ion{H}{$\alpha$} emission. On the other hand, biconical
models of the BLR seem needed to explain the variability of the rare
double-peaked AGNs (i.e. AGNs showing emission lines with two peaks)
and their polarisation properties (Sulentic et al.~\cite{SUL95},
Corbett et al.~\cite{COR98}). These studies show that there is no
consensus on the geometry of the BLR. From an observational and
theoretical perspective, it seems nevertheless to depend on the
ionisation degree of the line and on the radio properties of the
object.

\subsection{The radius-luminosity relationship}
\label{subsec:RL}

The ``size'' of the the broad line region $R_{BLR}$ has been measured
in about 40 AGNs by use of the reverberation mapping technique
(e.g. Krolik et al.~\cite{KRO91}, Horne et al.~\cite{HOR04}). The
empirical relation $R_{BLR} \propto L^\alpha$ ($\alpha\sim$0.5-0.6,
Kaspi et al.~\cite{KAS05}, Bentz et al.~\cite{BEN06}), combined with
the virial theorem, allows one to derive a relation linking the black
hole mass $M_{BH}$, the AGN luminosity $L$ and the FWHM of the
emission line. This relation is one of the most popular methods used to
measure black hole masses based on single epoch spectroscopic data and
study their growth, evolution and correlation to other AGN properties.

The $R_{BLR} \propto L^\alpha$ relation has been derived quite
accurately for the broad component of the \ion{H}{$\beta$} line, but
only a few objects have $R_{BLR}$ measurements for high ionisation
lines like \ion{C}{IV} (e.g. Kaspi et al. ~\cite{KAS07}). This is a
severe problem for black hole mass measurements of high redshift
objects.  The use of the \ion{C}{IV} line to derive black hole masses
is desirable but faces several problems related to our understanding
of the structure and geometry of the region emitting this line (see
e.g. Marziani et al. ~\cite{MAR06} for a review). The existence and
contamination of a narrow emission component in \ion{C}{IV} which
could bias FWHM$_{\ion{C}{IV}}$ (Bachev et al.~\cite{BAC04}) and the
possible absence of virial equilibrium, especially in sources showing
large blueshifts (Richards et al.~\cite{RIC02}), are major concerns.

Velocity resolved reverberation mapping (Horne et al.~\cite{HOR04})
should allow one to get insights on these problems, but the technique
is still under development and has not yet been applied to the
\ion{C}{IV} line. Most of the recent advances in velocity resolved
echo-mapping are based on the analysis of the \ion{H}{$\beta$}
emission line (Denney et al.~\cite{DEN09}, Bentz et
al.~\cite{BEN09b},~\cite{BEN10}). Although the kinematics of the
Balmer gas was thought to be relatively simple, the technique provided
puzzling results as both keplerian rotation, inflow and outflow
signatures appear in various objects. It is still unclear whether this
reveals a wide variety of kinematics, a biased interpretation of the
observed signal, or a spurious signal introduced by observational
artifacts. The past evidence of inflow and outflow signatures in NGC
5548 (Clavel et al.~\cite{CLA91}, Peterson et al.~\cite{PET91},
Kollatschny \& Dietrich~\cite{KOL96}) might indicate that
inflow/outflow signatures are not unambiguous. A possible explanation
of the variety of observed signals might be off-axis illumination of
the BLR (Gaskell~\cite{GAS10a}, \cite{GAS10b} and references therein).

\subsection{Microlensing in \obj}

The many open questions concerning the BLR we outlined above motivate
the interest in developing new techniques of probing this region.  In
this paper we investigate the constraints on the broad lines provided
by means of the microlensing study of the lensed quasar \obj, for
which long-term spectro-photometric monitoring has been carried out
(Eigenbrod et al.~\cite{EIG07}, hereafter {\it{Paper I}}). The
gravitational lens \obj, also known as ``Huchra's lens'' or the
``Einstein Cross'', was discovered by Huchra et al.  (\cite{HUC85})
during the Center for Astrophysics Redshift Survey.  It consists of a
$z_s=1.695$ quasar gravitationally lensed into four images arranged in
a cross-like pattern around the nucleus of a $z_l=0.0394$ barred Sab
galaxy. The average projected distance of the images from the lens
centre is only 700\,pc $\sim$ 0.9\arcsec, such that the matter along
the line of sight to the lensed images is mostly composed of
stars. The symmetric configuration of the lensed images around the
lens-galaxy bulge ensures a time delay of less than a day between the
lensed images, such that intrinsic flux variations should be seen
quasi- simultaneously in the four images (Rix et al.~\cite{RIX92},
Wambsganss \& Paczy\`nski~\cite{WAM94}). Since the low redshift of the
lens galaxy leads to a high relative transverse velocity between the
observer, the lens, and the source, the lensed images of \obj\, are
continuously flickering due to microlensing produced by the stars in
the lens galaxy, on smaller timescales than in any other lens. The
microlensing affecting the images of \obj\, leads to large variations
in amplitude (Udalski et al.~\cite{UDA06}), reaching up $>$ 1 mag,
often accompanied by chromatic microlensing of the quasar continuum
(Wambsganss \& Paczy\,nski~\cite{WAM91}) used to study the accretion
disc temperature profile (e.g. Kochanek~\cite{KOC04}, Anguita et
al.~\cite{ANG08}, Eigenbrod et al.~\cite{EIG08}={\it Paper
  II}). Several studies have shown that the broad emission lines are
also significantly affected by microlensing (Metcalf et
al.~\cite{MET04}, Wayth et al.~\cite{WAY05}, Paper I). Based on the
microlensing of \ion{C}{III]} observed at one epoch in 2002, Wayth et
al. (\cite{WAY05}) derived a most likely size of the region emitting
this line of 0.06 $h^{1/2}$ pc. The work presented here aims at
improving the measurement of the BLR size by using monitoring data
instead of a single epoch measurement and at constraining the
structure of the BLR.

The structure of the paper is the following. In Sect.~\ref{sec:obs},
we briefly summarise the data we used and explain the three different
methods we applied to analyse the spectra. In
Sect.~\ref{sec:phenomenology}, we apply these techniques to our data
to understand how microlensing deforms the emission lines and to
measure the flux ratios in several portions of the \ion{C}{IV} and
\ion{C}{III]} emission lines. In Sect.~\ref{sec:simulation}, we
explain the microlensing simulations we developed to derive the size
of the broad emission lines and continuum emission region. In
Sect.~\ref{sec:results} we present our results: measurement of the
size of the BLR and of the continuum region, comparison of our BLR
size with reverberation mapping, estimate of the black hole mass, and
finally constraints on the BLR structure. Finally, we summarise our
main findings in Sect.~\ref{sec:conclusions}.

\section{Observations and analysis techniques}
\label{sec:obs}

We compare emission line fluxes in images A \& D of \obj. A
comprehensive description of the observations and of the observational
setup can be found in Papers I \& II. The full data set consists of a
series of spectra obtained at 39 different epochs between
October 2004 and December 2007 with the FORS1 instrument at the ESO
Very Large Telescope. All the observations were carried out in MOS
(Multi Object Spectroscopy) mode. The multi-object mask used to obtain
the data presented here included the quasar images A and D, as well as
several field stars which are used for simultaneous flux calibration
and deconvolution. The spectra of the individual lensed quasar images
were de-blended (from each other and from the lensing galaxy) using
the MCS deconvolution technique (Magain et al.~\cite{MAG98}). A slit
width of 0.6\arcsec was used together with grism GRIS\_300V and
blocking filter GG375, leading to a resolving power $R=400$ at
5900\,$\AA$. For the data obtained in 2007, the FORS CCD had been
replaced by a new camera that was more sensitive in the blue
wavelength range. With this new camera, the GG375+30 blocking filter
was not used. The spectral resolution was improved by about 10\%, but
the observations obtained with the new setup have a signal-to-noise
ratio S/N$\lesssim$10 in the red wavelength range ($\lambda >
6700$\,\AA) due to fringing.  Table~\ref{tab:journal} provides a log of
the data sample we used. Data flagged with a $\dagger$ are
systematically removed as the \ion{C}{IV} emission is not covered by
our spectra.

\begin{figure}[t!]
\begin{center}
\includegraphics[height=7.0cm]{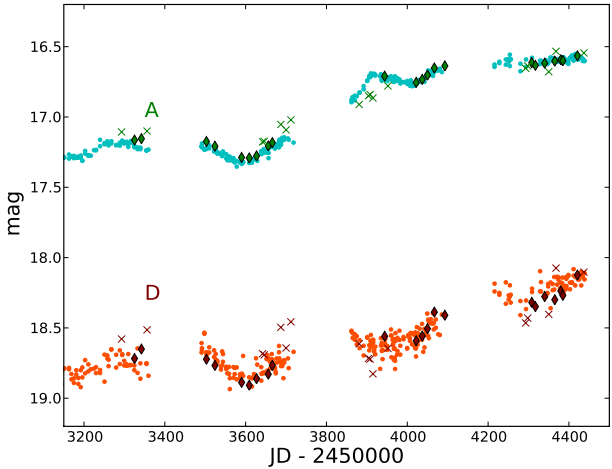}
\caption{Comparison of V-band OGLE lightcurves of images A \& D (points) of \obj\, with the lightcurves derived from our spectra (diamonds; crosses are used for epochs deviating from the OGLE lightcurves). Error bars (not displayed) are typically of the order of 0.02 mag (A) and 0.04 mag (D).}
\label{fig:oglecompa}
\end{center}
\end{figure}

We show in Fig.~\ref{fig:oglecompa} a comparison of the V-band
lightcurves of images A \& D provided by the OGLE collaboration
(Udalski et al.~\cite{UDA06}) with the lightcurves derived by applying
V-band response curve to the spectro-photometric monitoring
data. Throughout this paper, we shift the magnitude of image D from
the OGLE-III lightcurve by $-0.5$~mag with respect to the published
values to account for different flux calibration of that image in the
OGLE-III data (see Paper I for discussion). Although the overall
agreement is very good, several epochs identified with an 'X' symbol
deviate from the OGLE lightcurve. These data are flagged with an
asterisk ($*$) in Table~\ref{tab:journal} and we test how they affect
our results in Sect.~\ref{subsec:size}.


To study how microlensing deforms our spectra and how the broad lines
are affected, we use three different techniques that are explained
hereafter. First, we apply the multi-component decomposition
technique (MCD); i.e., we assume that multiple components cause the quasar
emission and measure the flux in these individual components
(Sect.~\ref{subsec:MCD}). Second, we use the macro-micro
decomposition technique (MmD); i.e., we follow the prescription of
Sluse et al. (\cite{SLU07}) and Hutsem\'ekers et al. (\cite{HUT10}) to
decompose the quasar spectrum into a component $F_M$, unaffected by
microlensing and another one, $F_{M\mu}$, which is microlensed
(Sect.~\ref{subsec:FMFMmu}). Third, we use the narrow band technique
(NBD); i.e., we cut the emission lines into different velocity slices
and measure the flux in each slice to look for microlensing within the
emission line (Sect.~\ref{subsec:bandemission}).

\begin{table}[tb]
\caption[]{Journal of the observations used in this paper. The observation ID refers to the whole data set presented in Papers I and II. Data flagged with a $\dagger$ are not used. Data flagged with an $*$ are those deviating from the OGLE lightcurve in Fig.~\ref{fig:oglecompa}. As in Paper II, epochs \#7, 8, 10, 30 have been excluded from the analysis. }
\label{tab:journal}
\begin{flushleft}
\begin{tabular}{lllcccc}
\hline 
\hline 
& ID & Civil Date & JD-2450000 & Seeing $[\arcsec]$ & Airmass \\
\hline 
\multirow{16}{*}{\rotatebox{90}{\mbox{Period P1}}} 					       
&1$^*$  & 2004$-$10$-$13& 3292 & 0.86 & 1.204 \\    
&2  & 2004$-$11$-$14& 3324 & 0.75 & 1.184 \\    
&3  & 2004$-$12$-$01& 3341 & 0.88 & 1.355 \\    
&4$^*$   & 2004$-$12$-$15& 3355 & 0.99 & 1.712 \\    
&5  & 2005$-$05$-$11& 3502 & 0.87 & 1.568 \\    
&6  & 2005$-$06$-$01& 3523 & 0.63 & 1.342 \\    
&9  & 2005$-$08$-$06& 3589 & 0.51 & 1.135 \\    
&11 & 2005$-$08$-$25& 3608 & 0.49 & 1.261 \\    
&12 & 2005$-$09$-$12& 3626 & 0.70 & 1.535 \\    
&13$^*$ & 2005$-$09$-$27& 3641 & 0.92 & 1.480 \\    
&14$^*$ & 2005$-$10$-$01& 3645 & 0.78 & 1.281 \\    
&15 & 2005$-$10$-$11& 3655 & 0.57 & 1.140 \\    
&16$^*$ & 2005$-$10$-$21& 3665 & 0.70 & 1.215 \\    
&17$^*$ & 2005$-$11$-$11& 3686 & 0.90 & 1.137 \\    
&18$^*$ & 2005$-$11$-$24& 3699 & 0.78 & 1.265 \\    
&19$^*$ & 2005$-$12$-$06& 3711 & 1.10 & 1.720 \\    
\hline
\multirow{11}{*}{\rotatebox{90}{\mbox{Period P2}}} 
&20$^\dagger$ & 2006$-$05$-$24& 3880 & 0.87 & 1.709 \\    
&21$^\dagger$ & 2006$-$06$-$16& 3903 & 0.66 & 1.213 \\    
&22$^{\dagger\,*}$ & 2006$-$06$-$20& 3907 & 0.64 & 1.286 \\    
&23$^\dagger$ & 2006$-$06$-$27& 3914 & 0.41 & 1.145 \\    
&24 & 2006$-$07$-$27& 3944 & 0.74 & 1.316 \\    
&25$^*$ & 2006$-$08$-$03& 3951 & 0.73 & 1.246 \\    
&26 & 2006$-$10$-$13& 4022 & 0.59 & 1.176 \\    
&27 & 2006$-$10$-$28& 4037 & 0.57 & 1.148 \\    
&28 & 2006$-$11$-$10& 4050 & 0.89 & 1.515 \\    
&29 & 2006$-$11$-$27& 4067 & 0.87 & 1.255 \\    
&31 & 2006$-$12$-$22& 4092 & 0.80 & 2.018 \\    
\hline
\multirow{12}{*}{\rotatebox{90}{\mbox{Period P3}}}
&32$^*$ & 2007$-$07$-$10 & 4292 & 0.63 & 1.153\\
&33$^*$ & 2007$-$07$-$15 & 4297 & 0.57 &	1.158\\
&34 & 2007$-$07$-$25 & 4307 & 0.79 &	1.231\\
&35 & 2007$-$08$-$03 & 4316 & 0.68 &	1.412\\
&36 & 2007$-$08$-$27 & 4340 & 0.88 & 1.133 \\
&37$^*$ & 2007$-$09$-$06 & 4350 & 0.67 & 1.396 \\
&38 & 2007$-$09$-$20 & 4364 & 0.73 &	1.230\\
&39$^*$ & 2007$-$09$-$23 & 4367 & 1.23 & 1.530 \\
&40 & 2007$-$10$-$05 & 4379 & 0.59 &	1.153\\
&41 & 2007$-$10$-$10 & 4384 & 0.76 &	1.283\\
&42 & 2007$-$11$-$15 & 4420 & 1.07 & 1.189 \\
&43$^*$ & 2007$-$12$-$01 & 4436 & 1.00  & 1.502 \\ 
\hline  					        
\end{tabular}					        
\end{flushleft} 				        
\end{table}	

\subsection{The multi-component decomposition (MCD)}
\label{subsec:MCD}

Our most comprehensive analysis technique is a multi-component
spectral decomposition of the quasar spectrum (MCD). With this
technique, we decompose the quasar spectra into the components likely
associated with different emission regions (Dietrich et
al. \cite{DIE03}). As in Papers I \& II, the spectrum is modelled as a
sum of three components: i) a power-law for the quasar continuum
emission, ii) an empirical iron template, and iii) a sum of Gaussian
profiles to model the emission lines. The exact decomposition of each
spectrum differs in several respects from the one presented in Papers
I \& II:
\begin{itemize}

\item First, the quasar emission in image $D$ is corrected for dust
  reddening produced by the lensing galaxy using the differential
  extinction values derived in Paper I. We adopt a Cardelli
  (\cite{CAR89}) Milky-way like extinction law with (Rv, Av) = (3.1,
  0.2 mag).

\item Because the UV \ion{Fe}{II+III} template constructed by
  Vestergaard \& Wilkes (\cite{VES01}) from I Zw1 does not perfectly
  reproduce the \ion{Fe}{II+III} blended emission observed in \obj\,,
  we construct the iron template empirically. In order to take
  possible intrinsic flux variations and/or systematic errors into
  account, we proceed independently for data obtained in 2004-2005
  (period {\bf {P1}}), 2006 (period {\bf {P2}}), and 2007 (period {\bf
    {P3}}). This template is constructed iteratively based on the
  residuals of the fitting procedure. In a first step, we fit only a
  power-law to the spectra, excluding those regions where
  \ion{Fe}{II+III} is known to be significant (Vestergaard \& Wilkes
  ~\cite{VES01}) and where there are noticeable quasar broad emission
  lines. The subsequent fits are performed on the whole quasar
  spectrum (excluding atmospheric absorption windows), fixing the
  width and centroid of the quasar emission lines. A new fit is then
  performed using the residuals as the iron template. The procedure is
  repeated until a good fit of the whole quasar spectrum is obtained.

\item In order to better reproduce the detailed shape of the
  \ion{C}{IV} and \ion{C}{III]} line profiles, we refine their
  modelling compared to Paper I by decomposing each line profile into
  a sum of three Gaussian profiles. The \ion{C}{IV} line is composed
  of two emission (a narrow-component NC and a broad-component BC) and
  of one absorption component (AC). The faint 1600\,\AA\, blend of
  emission sometimes observed in AGNs, and caused mostly by
  \ion{He}{II}\,$\lambda 1640$, \ion{O}{III]}$\lambda 1663$ (Fine et
  al.~\cite{FIN10} and ref. therein), is included by construction in
  our \ion{Fe}{II+III} template.  To reproduce the \ion{C}{III]} line,
  we need to use three emission components of increasing width
  (abbreviated NC, BC, and VBC). With the exception of
  \ion{C}{III]}(VBC), the central wavelength $\lambda_c$ of the
  Gaussians are allowed to vary from epoch to epoch as is the
  amplitude of each component. Table~\ref{tab:linedec} provides for
  each image, the central wavelength $\lambda_c$ and the FWHM of the
  different components of the line. The average value and standard
  deviation between epochs are displayed. When the standard deviation
  is not displayed, it means that the quantity is fixed for all epochs
  (see Appendix~\ref{sec:emiline} for details). Figure
  ~\ref{fig:linedec} shows the \ion{C}{IV} and \ion{C}{III]} emission
  line on 2004$-$11$-$14 and the corresponding profile
  decomposition. As in Paper I, the other emission line profiles
  (e.g. \ion{Al}{III}, \ion{He}{II}, \ion{Si}{III]}, \ion{O}{III]},
  \ion{Mg}{II}) are fitted with single Gaussian profiles of fixed
  width.
\end{itemize} 

\begin{table}[t]
\caption[]{Decomposition of the \ion{C}{IV} and \ion{C}{III]} emission lines (Average value and scatter between epochs when the parameter is fitted). The central wavelength $\lambda_c$ is expressed in \AA~ and the FWHM in km\,s$^{-1}$.}
\label{tab:linedec}
\begin{flushleft}
\begin{tabular}{ccccc}
\hline 
\hline
ID & \multicolumn{2}{c}{A} & \multicolumn{2}{c}{D} \\ 
\hline
& $\lambda_c$ & FWHM & $\lambda_c$ & FWHM \\
\hline  
\multicolumn{5}{c}{\ion{C}{IV}}\\
\hline
BC &     1547.8 $\pm$  1.0  &  6332 $\pm$  404  &  1546.9 $\pm$  0.9  &  6128 $\pm$  557  \\  
AC &     1548.9 $\pm$  0.5  &  1153 $\pm$  190  &  1548.7 $\pm$  0.4  &  1186 $\pm$  253  \\  
NC &     1547.6 $\pm$  0.6  &  2580  &  1547.0 $\pm$  0.4  &  2580  \\                    
\hline
\multicolumn{5}{c}{\ion{C}{III]}}\\
\hline
BC &   1910.7 $\pm$  0.6  &  3400  &  1910.7 $\pm$  1.1  &  3400 \\                     
NC &   1907.3 $\pm$  0.4  &  1545  &  1907.5 $\pm$  0.3  &  1545  \\                    
VBC &  1910.9   &  8548  &  1910.9  &  8548  \\               

\hline  					        
\end{tabular}					        
\end{flushleft} 				        
\end{table}

The results of the MCD decomposition are shown and discussed in
Sect.~\ref{sec:flux}. One could suspect that residual flux from the
lensing galaxy is included in our empirical FeII template. We tested
this possibility by fitting a lens galaxy template (as retrieved from
the deconvolution of the 2D spectrum, see Paper I) as an independent
component. However, the quasar lightcurves derived in this case
strongly disagree with the OGLE lightcurves, indicating that residual
contamination of the spectra by the lensing galaxy is not a big
issue. Finally, we emphasise that the decomposition of the carbon
lines in multiple Gaussians of different widths is empirical and does
not necessarily isolate emission lines coming from physically
different regions. The adopted naming of the components is based on
their width, and is line dependent and internal to this paper. Because
of the different decompositions found for the \ion{C}{IV} and
\ion{C}{III]} lines, comparison between Gaussian components should be
performed with care, keeping the actual line widths in mind.

\subsection{Macro-micro decomposition (MmD)}
\label{subsec:FMFMmu}

This technique, introduced in Sluse et al. (\cite{SLU07}) and
Hutsem\'ekers et al. (\cite{HUT10}) and previously nicknamed
$F_M-F_{M\mu}$ decomposition, allows us to determine the fraction of
the quasar emission that is not affected by microlensing independently
of any modelling of the quasar spectra.  Because spectral differences
between lensed images only stem from microlensing (here we correct for
differential extinction prior to the decomposition), we can assume
that the observed spectrum $F_i$ of image $i$ is the superposition of
an emission $F_M$ that is only macro-lensed and of another component
$F_{M\mu}$, both macro- and {\it micro}-lensed. Using pairs of spectra
of different lensed images, it is then easy to extract both components
$F_M$ and $F_{M\mu}$.  Defining $M=M_A/M_D \,(>0)$ as the
macro-magnification ratio between image A and image D, and
$\mu=\mu_A/\mu_D$ as the microlensing ratio between image A and image
D, we have

\begin{equation}
\begin{array}{l}
F_A = M \times F_M + M \times \mu \times F_{M\mu}\\
F_D = F_M + F_{M\mu} \,.\\
\end{array}
\label{eq:decomp1}
\end{equation}
\noindent
These 2 equations can be rewritten as
\begin{equation}
\begin{array}{l}
F_M = \frac{F_A/M - \mu \times F_D}{1-\mu} \\
F_{M\mu} = \frac{F_D- F_A/M}{1-\mu}\,, \\
\end{array}
\label{eq:decomp2}
\end{equation}
\noindent
where $\mu$ must be chosen to satisfy the positivity constraint $F_M >
0$ and $F_{M\mu} >0$. For the macro-magnification factor $M$, we adopt
the ratio $M=M_A/M_D=1.0$ provided by the lens model of the Einstein
cross performed by Schmidt et al. (\cite{SCH98}). The microlensing
ratio $\mu(t, \lambda)$ depends on wavelength and time and can be
estimated for each epoch based on the ratio of the continuum emission
between A \& D (Sect.~\ref{subsec:MCD}). Our choice of the continuum
flux ratio to derive $\mu(t, \lambda)$ means that we implicitly assume
that the emission line and pseudo-continuum emission arise from
regions equally or less microlensed than the continuum, in agreement
with theoretical expectations. This assumption is also verified a
posteriori as our analysis confirms that the continuum emission is
also the most microlensed.

\subsection{Narrow band technique (NBD) }
\label{subsec:bandemission}

\begin{figure*}[t]
\begin{center}
\includegraphics[height=7.0cm]{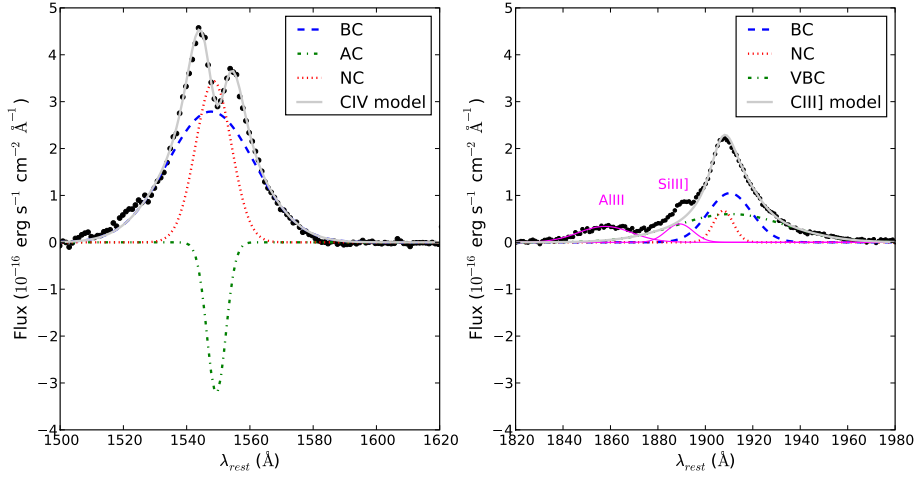}
\caption{Example of multi component decomposition of the \ion{C}{IV} ({\it{left}}) and \ion{C}{III]} ({\it {right}}) line profiles (Sect.~\ref{subsec:MCD}) after continuum and \ion{Fe}{II} subtraction. The narrowest component of the line (NC) is displayed with a dotted-red line, the broad component (BC) with a dashed-blue line, the very broad component (VBC) of \ion{C}{III]} and the absorber in front of \ion{C}{IV} are shown in green dot-dashed. The sum of the individual components, corresponding to the line model, is shown with a solid grey line. Characteristics of the components are provided in Table~\ref{tab:linedec}. }
\label{fig:linedec}
\end{center}
\end{figure*}

A simple method of identifying differential microlensing within an
emission line is to divide the line into various wavelength bands and
calculate the flux ratio $F_A/F_D$ for each of these bands.  For the
\ion{C}{IV} and \ion{C}{III]} emission, we chose three different
wavelength ranges corresponding to the blue wing (BE;
-8000 $<\,$v$\,< $ -1500 km\,s$^{-1}$), the line core (CE; -1500 $< \,$v$
\,< $ 1500 km\,s$^{-1}$), and the red wing (RE; 1500 $<\,$v$\, < $ 8000
km\,s$^{-1}$) of the line. Before integrating the flux in these ranges,
we first remove the continuum and the \ion{Fe}{II+III} emission from
the spectra based on the results of the MCD fit
(Sect.~\ref{subsec:MCD}). For the \ion{C}{III]} emission, we also
remove the \ion{Al}{III} and \ion{Si}{III]} emission. The central
wavelength corresponding to zero velocity is 1549\,\AA~for \ion{C}{IV}
and 1907.5\,\AA~for \ion{C}{III]}.

\section{Microlensing variability in the emission lines}
\label{sec:phenomenology}

\begin{figure*}[tb]
\begin{center}
\begin{tabular}{cc}
\includegraphics[width=9.0cm]{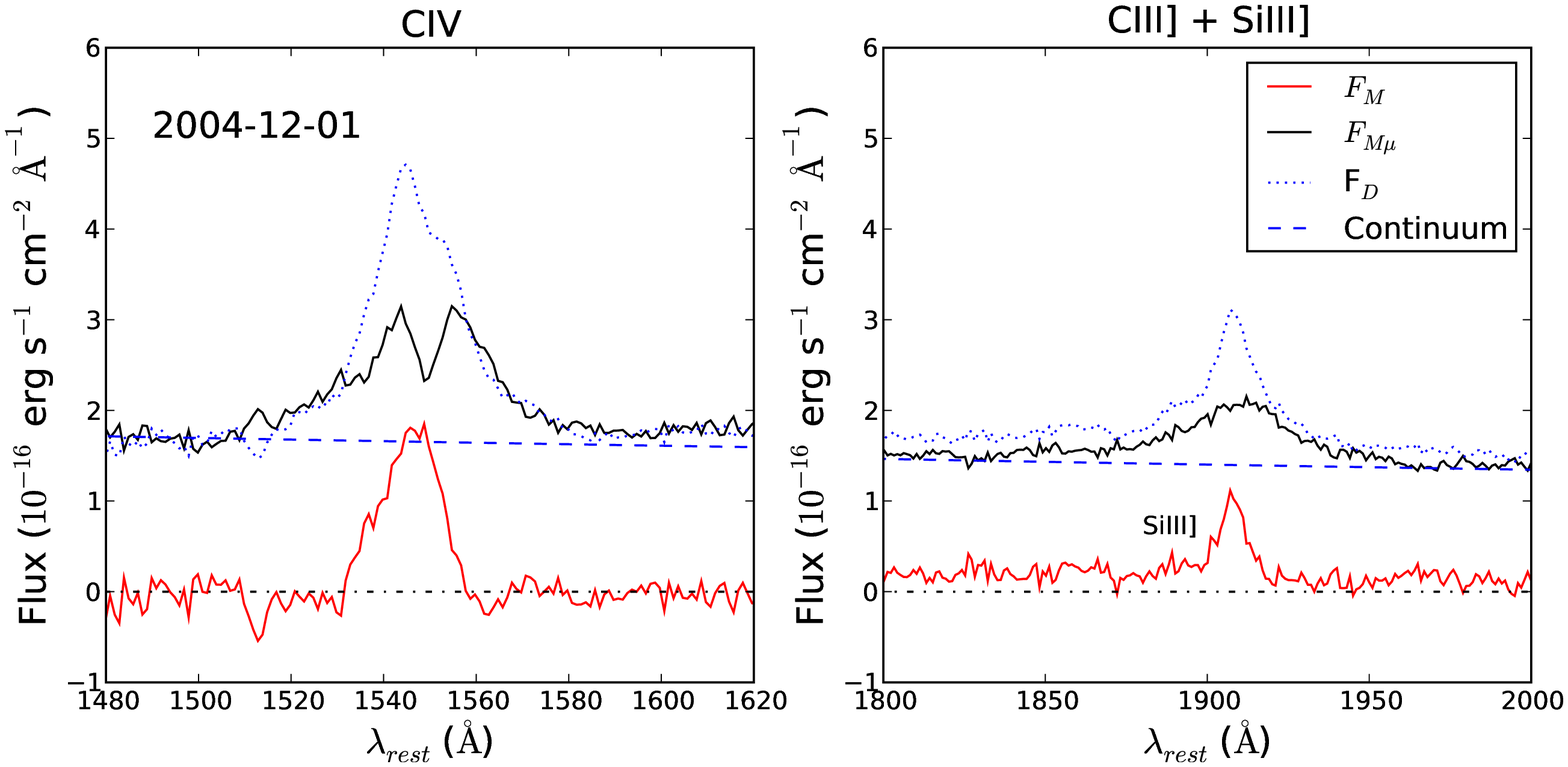}& \includegraphics[width=9.0cm]{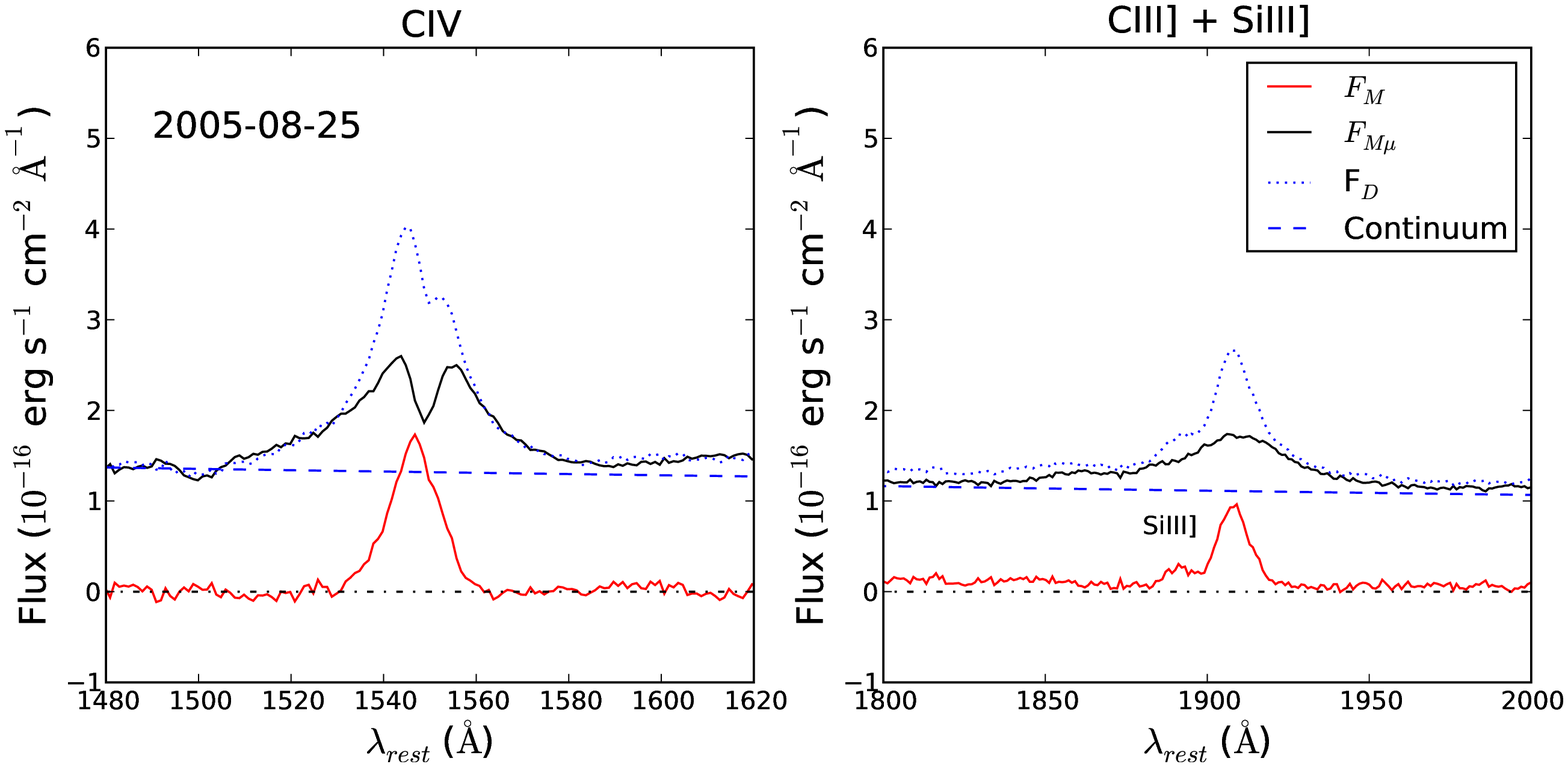}\\
\includegraphics[width=9.0cm]{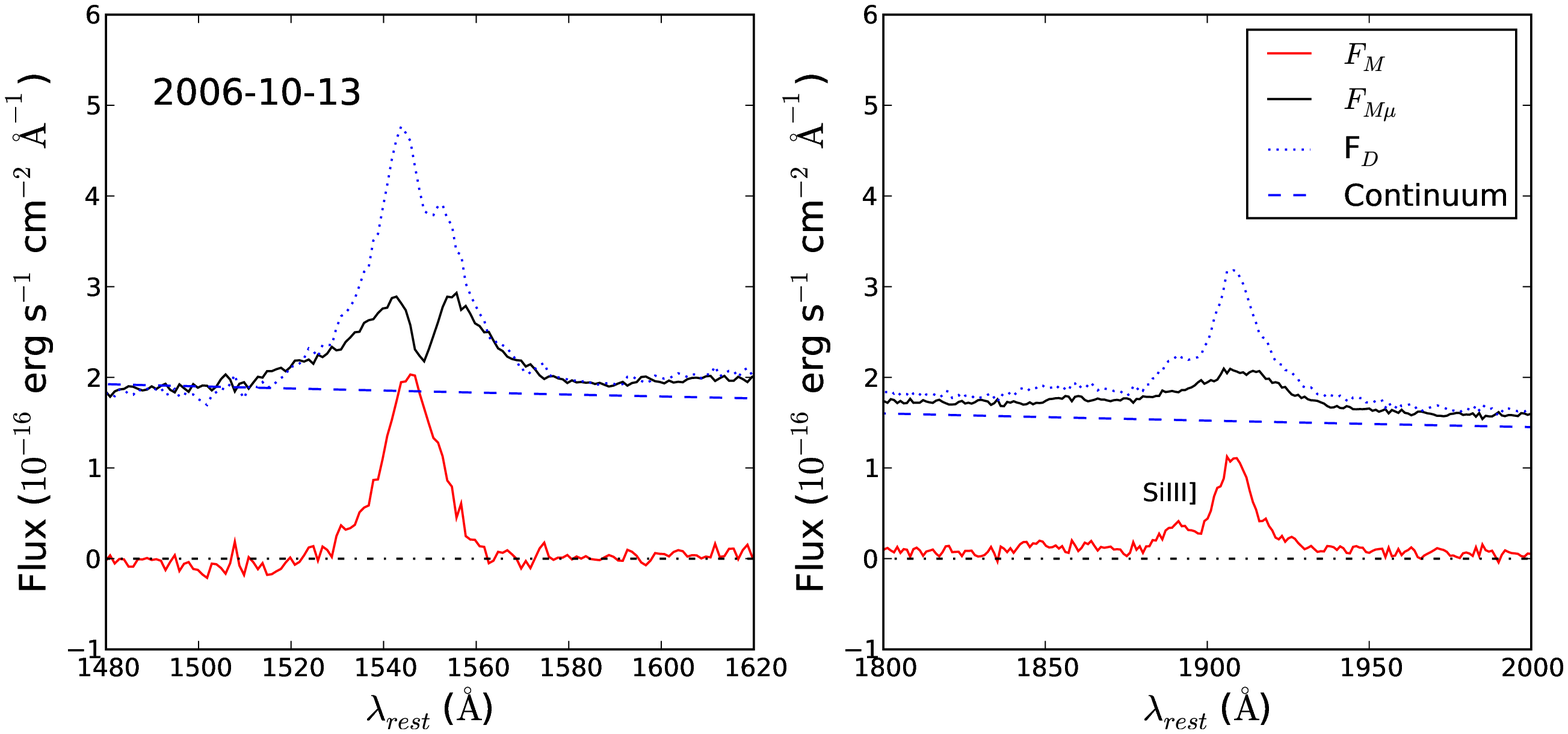}&\includegraphics[width=9.0cm]{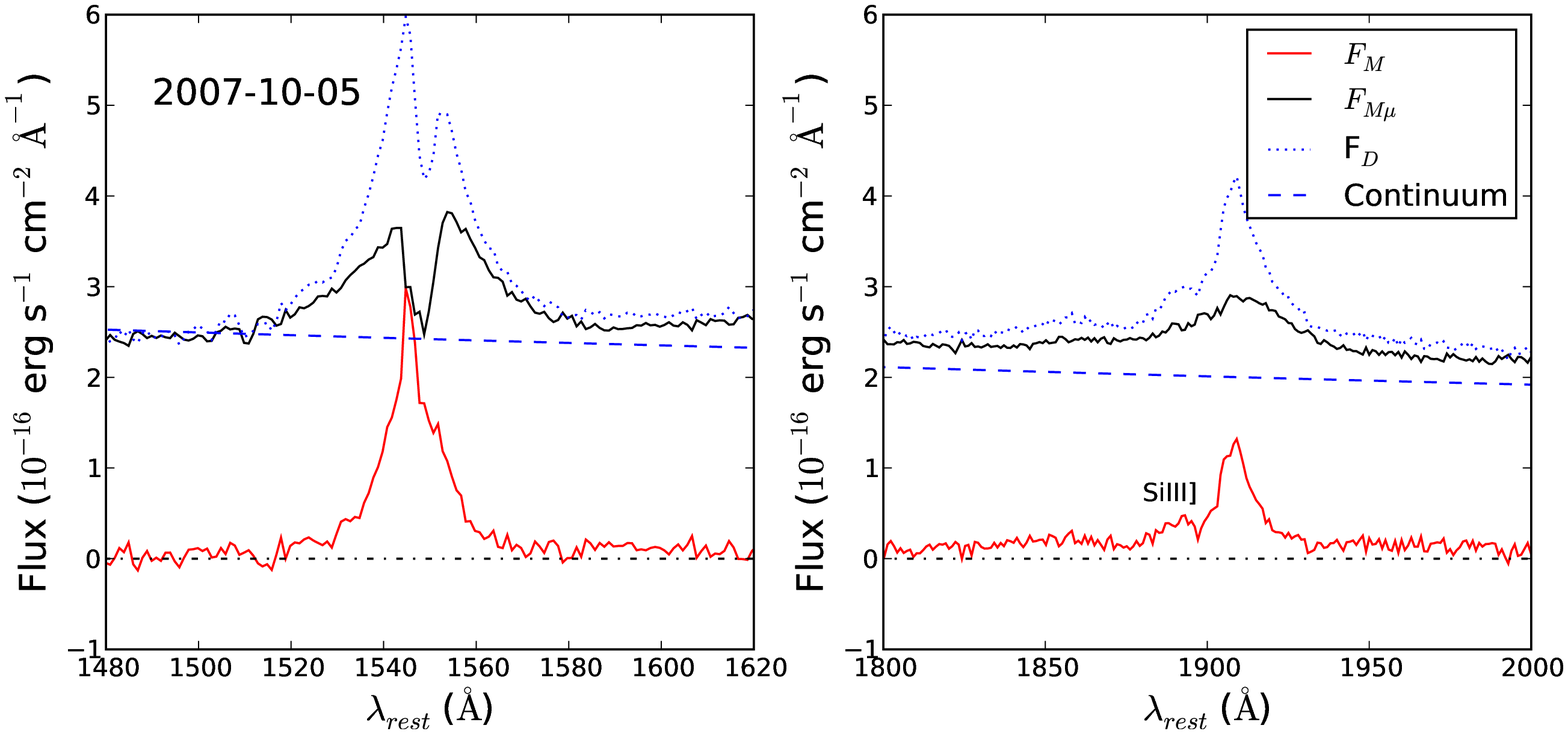}\\
\end{tabular}
\caption{Macro-micro decomposition (MmD) of the \ion{C}{IV} ({\it {left}}) and \ion{C}{III]}+\ion{Si}{III]} ({\it {right}}) emission lines obtained from the spectra of images A \& D at 4 different epochs. In each panel, the black solid line shows the fraction of the spectrum $F_{M\mu}$ affected by microlensing and the red solid line shows the emission $F_M$ which is too large to be microlensed. For comparison, we also display the observed spectrum of image D with a dotted blue line and the power-law continuum used to calculate $\mu(\lambda, t)$ with a dashed blue line.}
\label{fig:FMFMmu}
\end{center}
\end{figure*}

In Paper I, we reported evidence that the broad emission lines in
\obj\, are significantly microlensed in image A during the monitoring
period. In this section, we refine and extend these results (including
2007 data, which were absent from Paper I) using the techniques
outlined in Sect.~\ref{sec:obs}.  This will allow us to derive
lightcurves for the \ion{C}{IV} and \ion{C}{III]} emission lines and
identify signatures of microlensing.

\subsection{\ion{C}{IV} and \ion{C}{III]} emission lines}
\label{sec:flux}

\begin{figure*}[tb]
\begin{center}
\includegraphics[width=19.0cm]{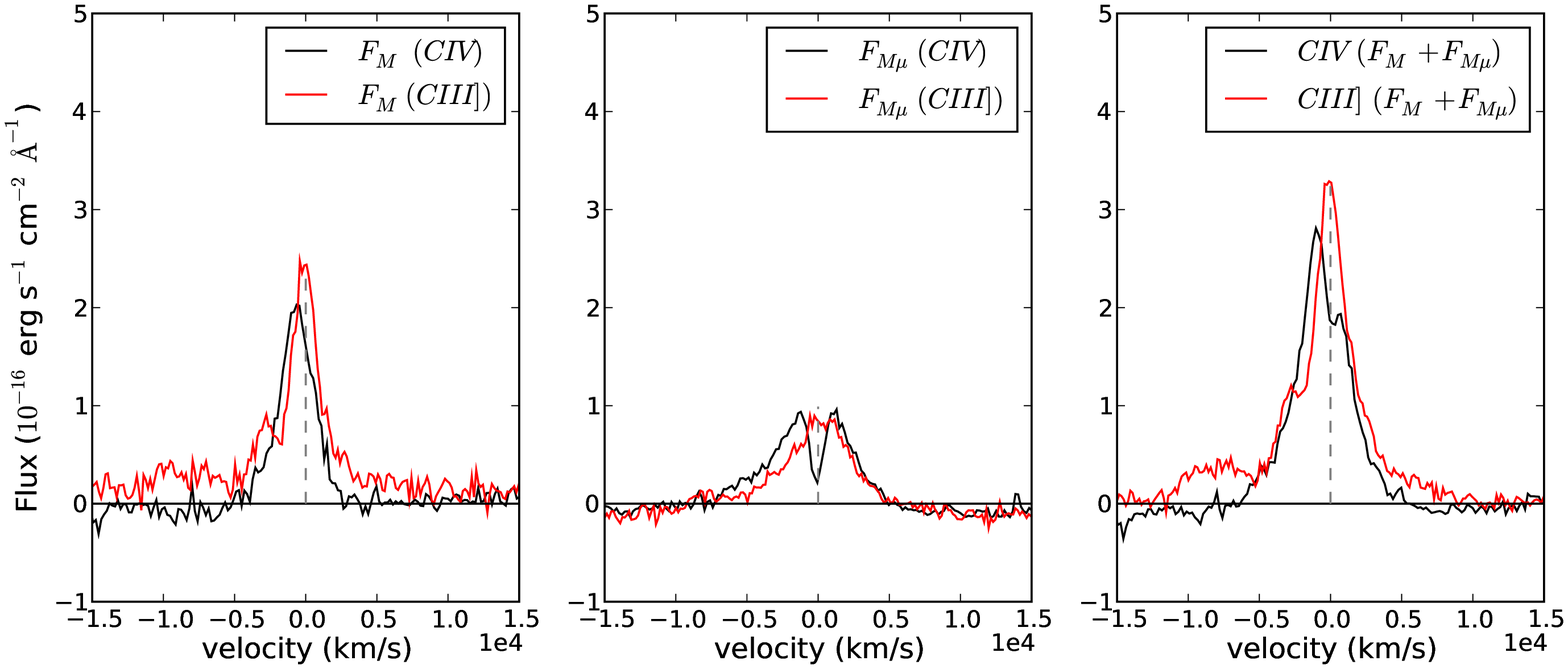}
\caption{Macro-micro decomposition (MmD) technique applied to the \ion{C}{IV} (black) and \ion{C}{III]} (red) emission lines observed on 2006-10-13. {\it{Left}}: Fraction of the flux $F_M$ non affected by microlensing, {\it{Centre}} Fraction of the line $F_{M\mu}$ affected by microlensing, {\it {Right:}} Full emission line profile. The intensity of \ion{C}{III]} are multiplied by 2.2 to ease visual comparison between \ion{C}{IV} and \ion{C}{III]} profiles. The vertical dashed line indicates the velocity zeropoint corresponding to the systemic redshift. }
\label{fig:velocity}
\end{center}
\end{figure*}

First, we investigate how microlensing affects the emission line
shape. For this purpose we use the MmD technique
(Sect.~\ref{subsec:FMFMmu}). The results of this decomposition for the
\ion{C}{IV} and \ion{C}{III]} emission are shown in
Fig.~\ref{fig:FMFMmu} for four representative epochs. In this figure,
we also overplot the continuum model derived from the MCD
decomposition and the spectrum of image D. As indicated
by Eq.~\ref{eq:decomp1}, the spectrum of image D should match
$F_{M\mu}$ in the region of the spectra which are microlensed like the
continuum ($F_{M\mu}$ = $F_D$ when $F_M=0$). The match between $F_D$
and $F_{M\mu}$ in the wings of the emission lines reveals that the
corresponding emitting region is microlensed as much as the continuum
(Figure~\ref{fig:FMFMmu}). It is also apparent that the microlensed
fraction of the lines has a broad Gaussian shape with no flux
depletion or flat core in the centre (the depletion observed for the
\ion{C}{IV} emission is caused by the intrinsic absorber). On the
other hand, in $F_M$, only a narrow emission line is visible for
\ion{C}{IV} and \ion{C}{III]}. If we remove the continuum power-law
model and \ion{Fe}{II} model prior to the decomposition, and plot the
\ion{C}{III]} and \ion{C}{IV} lines on a velocity scale
(Figure~\ref{fig:velocity}), we find that the decomposed line profiles
are rather similar. The remainder of the difference could be caused by
\ion{Fe}{II} blended with the \ion{C}{III]} emission, hence
not included in our \ion{Fe}{II} template, but may also be real. It is
also apparent that the line profile is asymmetric with a stronger blue
component and that the \ion{C}{IV} line is slightly blueshifted
($\sim$500km\,s$^{-1}$) compared to \ion{C}{III]}. Although comparison
between epochs is difficult due to intrinsic variability of the
emission lines, we observe that the decomposition remains roughly the
same during the period P1 (not shown). During periods P2 and P3, both
fractions $F_M$ and $F_{M\mu}$ of the carbon lines follow the
intrinsic brightening of the continuum.  We also tentatively see that,
during period P3, the line wings appear in $F_M$, indicating that the
broadest component is differently microlensed than the continuum at
these epochs.  We note that the use of a macro-magnification ratio $M
\sim $0.85, closer to the radio and mid-infrared flux ratios (Falco et
al.~\cite{FAL96}, Agol et al.~\cite{AGO09}, Minezaki et
al.~\cite{MIN09}), does not change the general features observed in
Figure~\ref{fig:FMFMmu}.

\begin{figure}[tbh]
\begin{center}
\begin{tabular}{c}
\includegraphics[width=8.0cm]{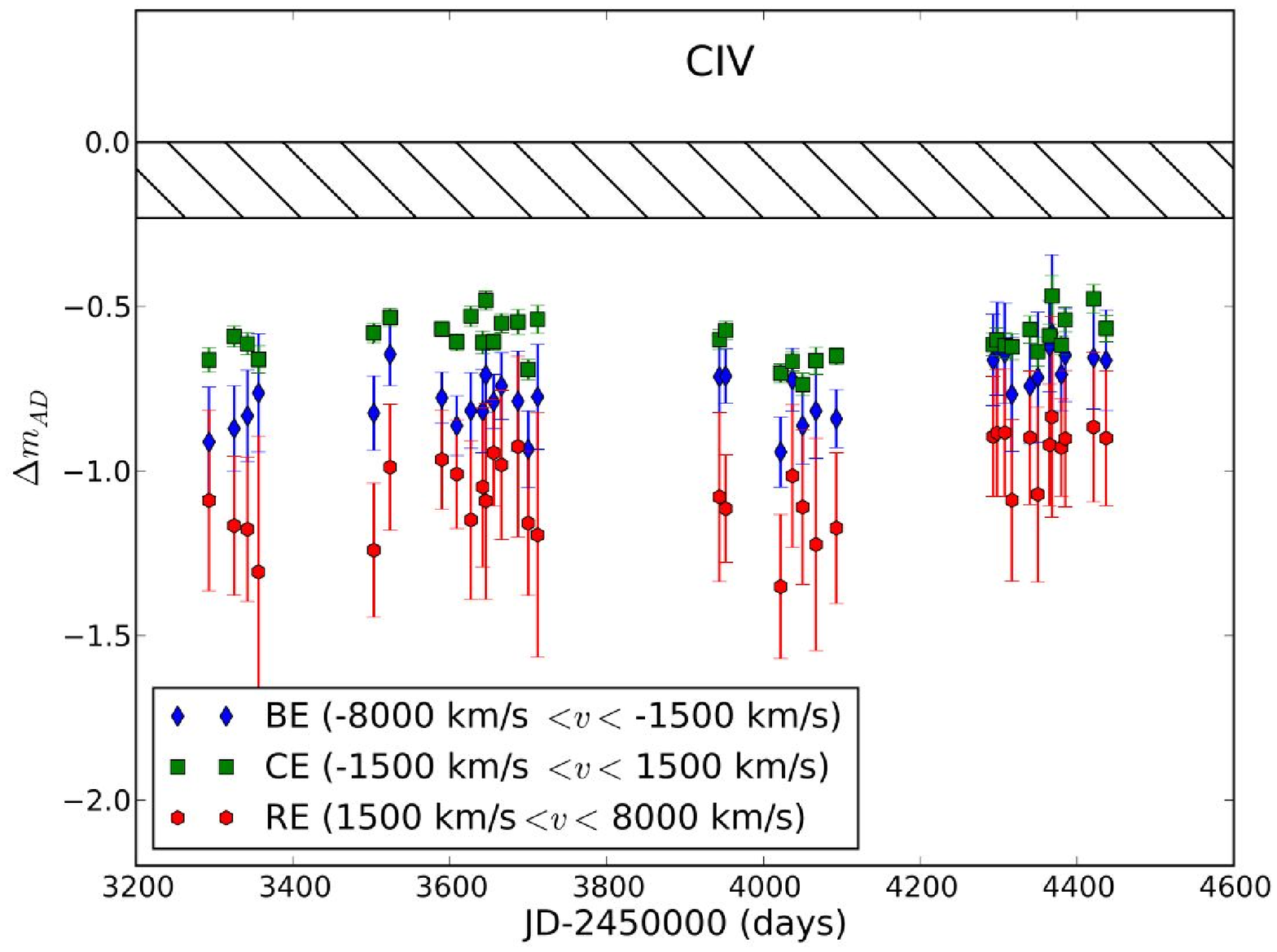} \\
\includegraphics[width=8.0cm]{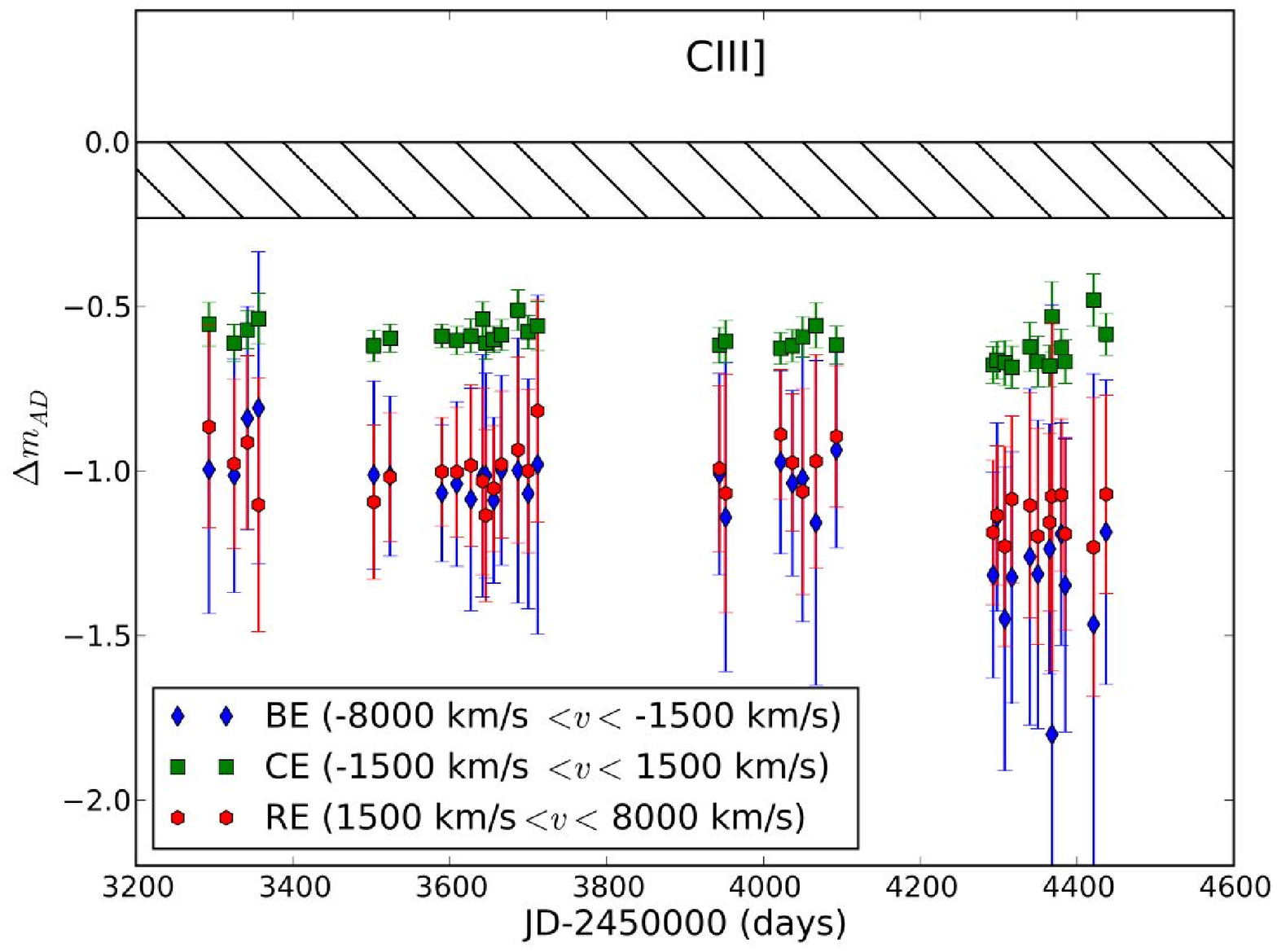}\\
\end{tabular}
\caption{Time dependence of the magnitude difference between images A \& D, $\Delta m_{AD}$, as measured in the blue wing (BE; blue diamond), line core (CE; green square) and red wing (RE; red hexagons) of the \ion{C}{IV} ({\it {Top}}) and \ion{C}{III]} ({\it {bottom}}) emission lines.  The shaded area indicates the range of $\Delta m_{AD}$ estimated from the macro-model and from MIR measurements.}
\label{fig:narrowband}
\end{center}
\end{figure}

The slicing of the emission lines in velocity also allows us to study
differential microlensing in the emission lines independently of a
model of the line profile.  Figure~\ref{fig:narrowband} displays the
flux measured in the blue wing (BE), red wing (RE), and line core (CE)
of the carbon emission lines, using the NBD
(Sect.~\ref{subsec:bandemission}). Standard error bars are calculated
based on the photon noise. Although the error bars are large, we
clearly see a difference of about 0.3 mag between the flux ratios
measured in the core and the wings of the emission lines.
Table~\ref{tab:bandratios} shows the average flux ratios for periods
P1, P2, P3. Error bars for a period are standard errors calculated
based on the dispersion of the data points. The systematic difference
observed between the blue and red wings of the \ion{C}{IV} profile
seems to be mostly caused by an asymmetry of the line's core. Indeed,
the offset between blue and red wing decreases as we increase the
width of the central velocity range (not shown). The asymmetry of the
$F_M$ fraction of the line visible in Figure~\ref{fig:velocity}
supports this finding. When looking to the intrinsic variability of
the different bands (i.e. individual lightcurves for images A \& D),
we observe that the wings vary more significantly on short timescales
than does the core of the line.

\begin{figure*}[tb]
\begin{center}
\begin{tabular}{ccc}
%
\includegraphics[width=5.5cm,   bb = 0 170 550 600]{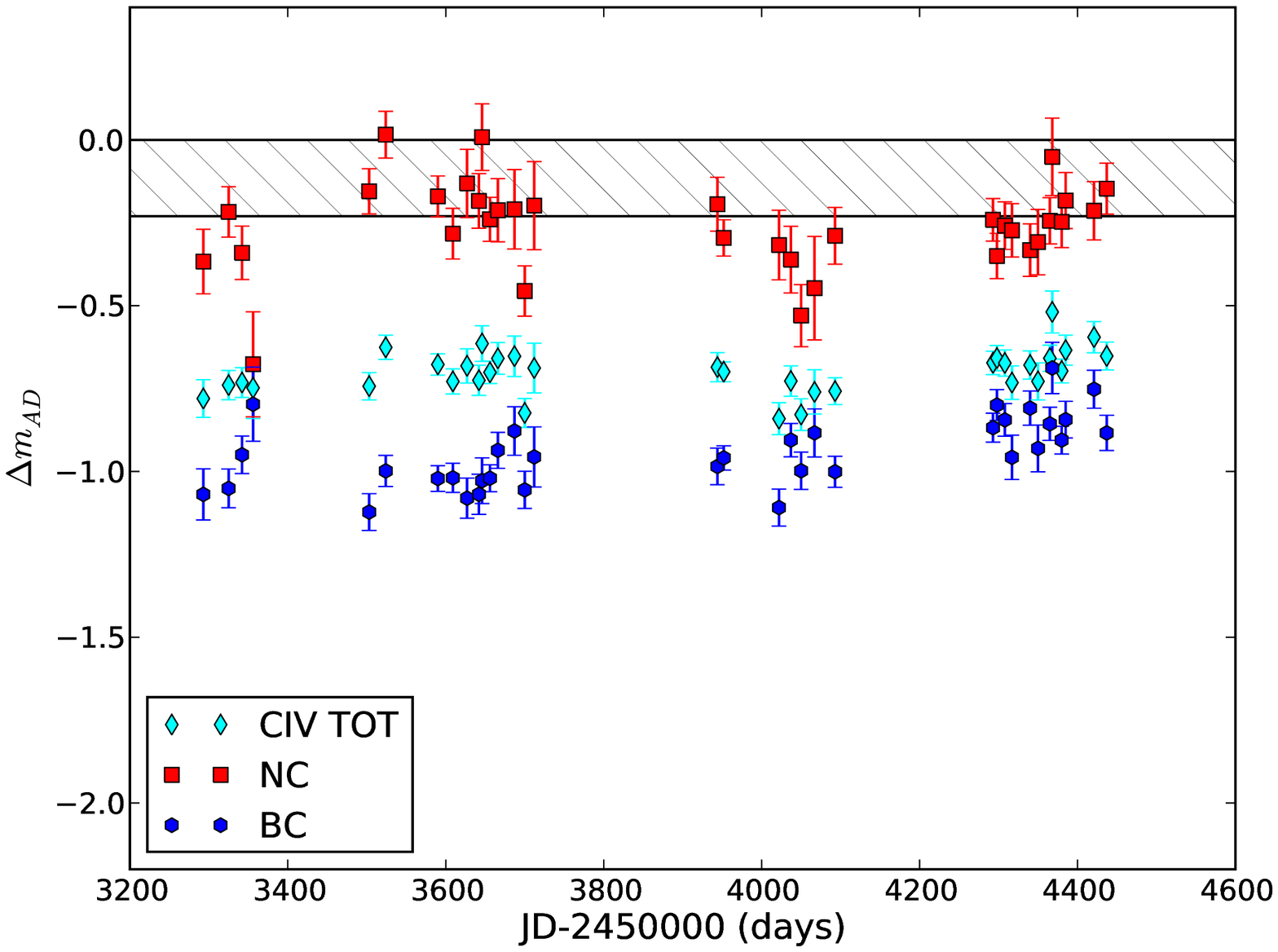}  &  \includegraphics[width=5.5cm,   bb = 0 170 550 600]{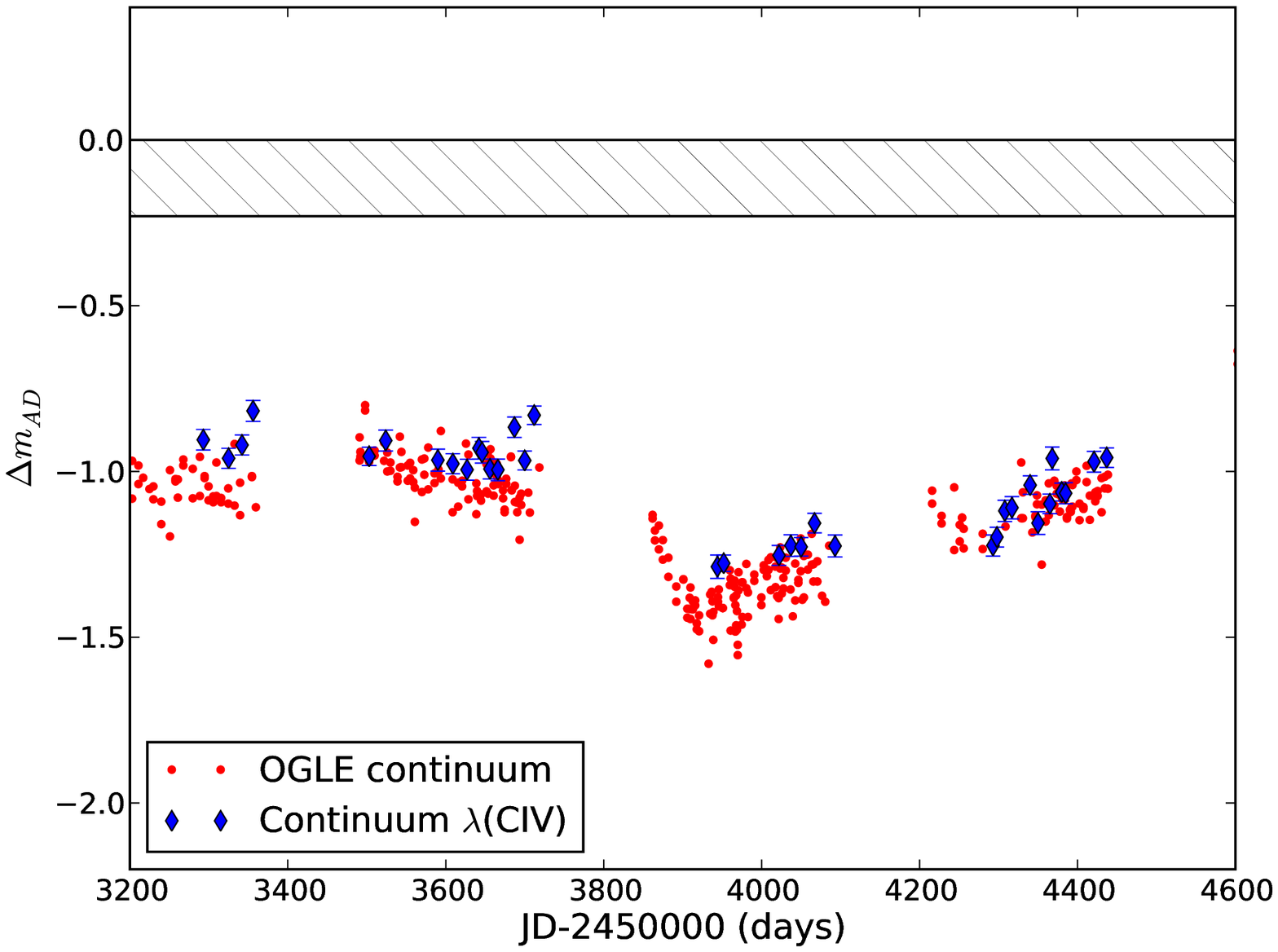}  & \includegraphics[width=5.5cm,  bb = 0 170 550 600]{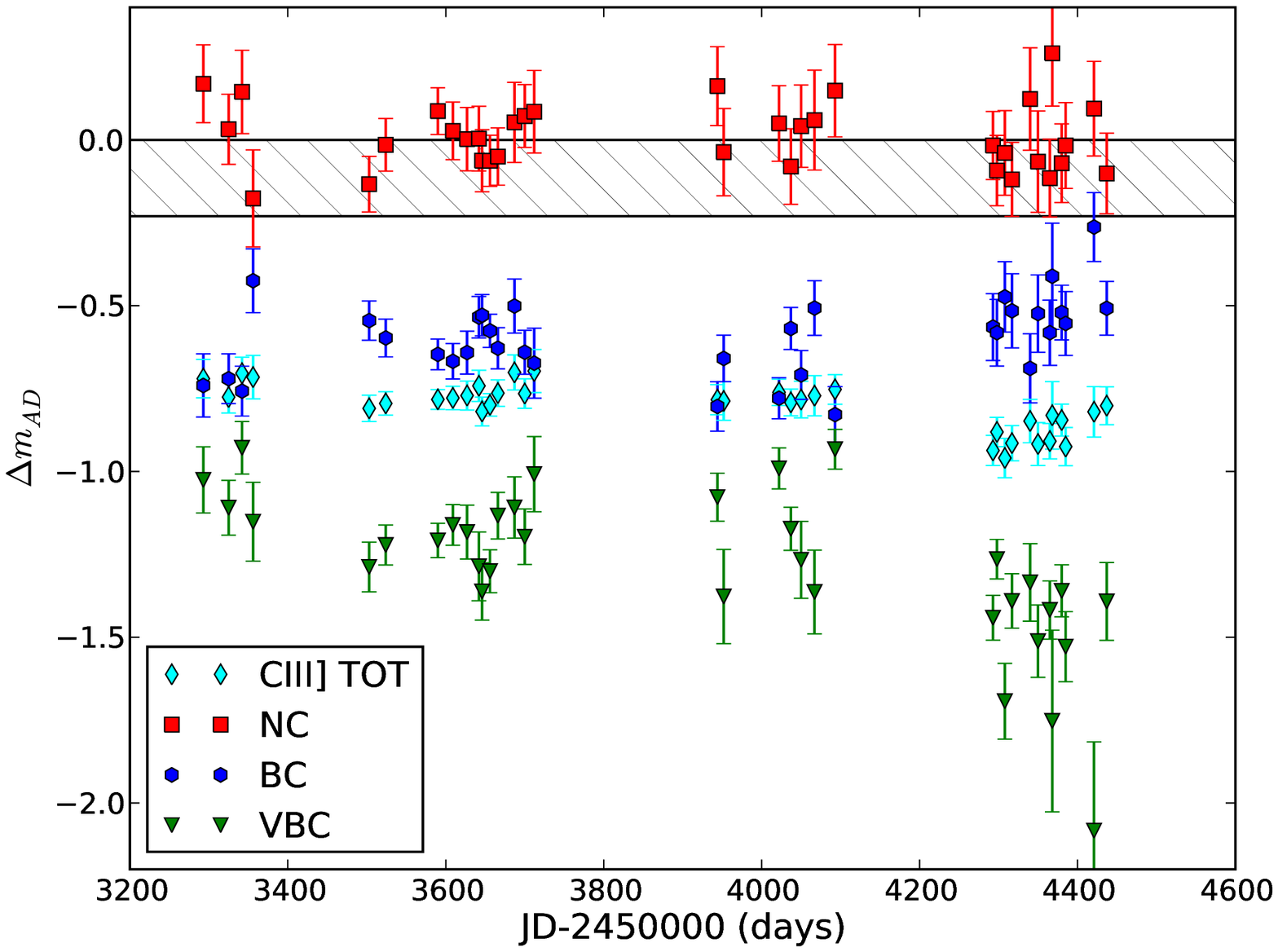} \\
\end{tabular}
\caption{Time variation of the magnitude difference between images A \& D, $\Delta m_{AD}$, as measured in the continuum ({\it {centre}}) and in the different components (Fig.~\ref{fig:linedec}) of the  \ion{C}{IV} ({\it {left}}), \ion{C}{III]} ({\it {right}}) emission lines. The shaded area indicates the range of $\Delta m_{AD}$ estimated from the macro-model and from MIR measurements. The acronyms NC, BC and VBC refer to the different components (of increasing width) of the \ion{C}{IV} \& \ion{C}{III]} lines as shown in Fig.~\ref{fig:linedec}.}
\label{fig:MCD}
\end{center}
\end{figure*}

\begin{table}[tbh]
\caption[]{Average values of magnitude difference $\Delta m_{AD}$ as measured in the blue wing (BE), line core (CE), and red wing (RE) of the \ion{C}{IV} and \ion{C}{III]} emission lines. }
\label{tab:bandratios}
\begin{center}
\begin{tabular}{ccccc}
\hline 
\hline 
& & \multicolumn{3}{c}{ $\Delta m_{AD}$ (mag)}\\
\hline
Line & Period & BE & CE & RE \\ 
\hline  

\ion{C}{IV} & P1&-0.91 $\pm$ 0.02 & -0.66 $\pm$ 0.02 & -1.27 $\pm$ 0.02 \\
 & P2&-0.80 $\pm$  0.03& -0.66 $\pm$ 0.02& -1.21 $\pm$ 0.04 \\
 & P3&-0.66 $\pm$  0.02& -0.58 $\pm$ 0.02 & -1.10 $\pm$ 0.02 \\
\ion{C}{III]} & P1&-1.08 $\pm$ 0.02 & -0.59 $\pm$ 0.01 & -1.07 $\pm$ 0.02 \\
& P2&-1.18 $\pm$  0.03& -0.62 $\pm$ 0.01& -1.14 $\pm$ 0.03 \\
& P3&-1.32 $\pm$  0.05& -0.61 $\pm$ 0.02 & -1.13 $\pm$ 0.02 \\

\hline  					        
\end{tabular}					        
\end{center} 				        
\end{table}

The results of the MCD decomposition are presented in
Figure~\ref{fig:MCD}. They broadly agree with the velocity slicing
technique (Fig.~\ref{fig:narrowband},
Table~\ref{tab:bandratios}). Figure~\ref{fig:MCD} shows that the flux
ratio in the full \ion{C}{III]} profile varies weakly over
the monitoring interval. A similar flux ratio is measured in
\ion{C}{IV} and \ion{C}{III]} with evidence of a small increase
($\Delta m_{AD}$ closer to 0) in 2007. Figure~\ref{fig:MCD} also
confirms that the flux ratio measured in the narrower component of the
emission lines is closer to 1 (i.e. the macro-lensed ratio $M_A/M_D$),
while the flux ratio amounts to nearly 2.5 in the broadest component
of the lines (i.e. BC of \ion{C}{IV} and VBC of \ion{C}{III]}).

\subsection{Other emission lines}
\label{sec:otherlines}

\begin{figure}[tbh]
\begin{center}
\includegraphics[width=9.0cm]{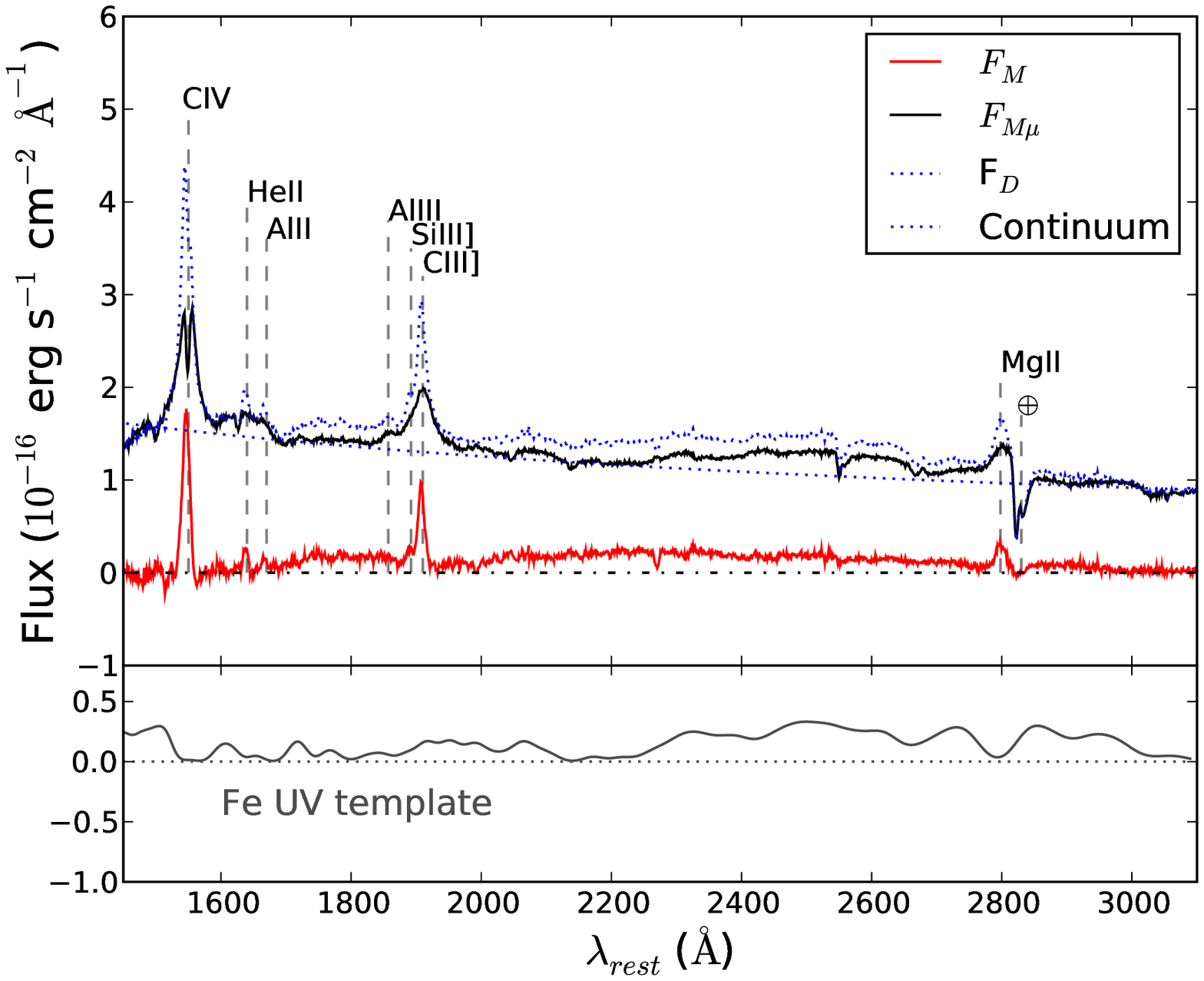} 

\caption{{\it {Top}}: Macro-micro decomposition (MmD) applied to the
  spectra of images A \& D averaged over period P1. The black solid
  line shows the fraction of the spectrum $F_{M\mu}$ affected by
  microlensing, and the red solid line shows the emission $F_M$, which
  is too large to be microlensed. For comparison, we also display the
  observed spectrum of image D with a dotted-blue line and the
  power-law continuum of D with a dashed black line. {\it{Bottom}}:
  \ion{Fe}{II+III} template from Vestergaard \& Wilkes~\cite{VES01}
  convolved with a Gaussian of FWHM$=$2000 km\,s$^{-1}$. }

\label{fig:FMFMmu2}
\end{center}
\end{figure}

An accurate study of the microlensing occurring in the other emission
lines is very difficult because of the pseudo-continuum
\ion{Fe}{II+III} in emission blended with these lines. The MmD
technique gives qualitative information about microlensing for the
other emission lines. Figure~\ref{fig:FMFMmu2} shows the $F_M$ and
$F_{M\mu}$ spectra averaged over period P1. We clearly see that not
only do \ion{C}{IV} and \ion{C}{III]} have their broadest component
microlensed, but also \ion{Mg}{II}\,$\lambda 2798$,
\ion{Al}{II}\,$\lambda 1671$, \ion{He}{II}\,$\lambda 1640$,
and \ion{Al}{III}\,$\lambda 1857$. The narrow component of these lines is
clearly visible in $F_M$. Interestingly, the \ion{Si}{III]}\,$\lambda
1892$ is not microlensed. The absence of microlensed \ion{Si}{III]}
emission is also clear in Fig.~\ref{fig:FMFMmu}. The bottom
panel shows the Fe$_{UV}$ template of Vestergaard \& Wilkes
(\cite{VES01}) convolved with a Gaussian of FWHM $\sim$ 2000
km\,s$^{-1}$. Although the iron emission in \obj\, differs slightly
from the Vestergaard \& Wilkes (\cite{VES01}) template (Paper I,
Sect.~\ref{subsec:MCD}), we clearly see that only the microlensed
pseudo-continuum (i.e. emission above the continuum in F$_{M\mu}$)
resembles the Fe$_{UV}$ emission but, there is no sign of UV iron
emission in $F_M$. The pseudo-continuum emission visible in $F_M$ at
$\lambda >$ 1920\,\AA~ is unlikely to be produced by Fe$_{UV}$ because
its shape deviates significantly from expectations. The most likely
origin(s) of this emission is (are) UV emission from the host galaxy,
Balmer continuum emission, and scattered continuum light. Although this
feature peaks at about $2200$\,\AA\, in the quasar rest frame, it is
unlikely to be caused by differential extinction in the quasar host
galaxy because the light-rays separation between A \& D in the host is
too small (Falco et al.~\cite{FAL99}).

\section{Microlensing simulations}
\label{sec:simulation}

In order to derive the size of the regions emitting the carbon lines,
we must compare the observed microlensing signal with simulated
microlensing lightcurves generated for different source sizes. The
strategy we adopt is similar to the one explained in Paper II. Instead
of searching for the best simulated lightcurve reproducing the data,
we follow a Bayesian scheme for which a probability is associated to
each simulated lightcurve. A probability distribution is then derived
for each parameter that describes the lightcurve. We provide a
description of the crucial steps in the next sections.

\subsection{Creating the microlensing pattern}
\label{subsec:pattern}

We use the state-of-the-art inverse ray-shooting code developed by
Wambsganss (\cite{WAM90}, \cite{WAM99}, \cite{WAM01}) to construct
micro-magnification maps for images A \& D. To create these patterns,
the code follows a large number of light-rays (of the order of
10$^{10}$), from the observer to the source plane, through a field of
randomly distributed stars. The surface density of stars and the shear
$\gamma$ at the location of the lensed images are those provided by a
macro model of the system. Instead of the surface density, we used the
dimensionless quantity $\kappa$ (the convergence), defined as the
ratio between the surface density and the critical density.  We use
($\kappa$, $\gamma$)$_A$ = (0.394, 0.395) and ($\kappa$, $\gamma$)$_D$
= (0.635, 0.623) (Kochanek~\cite{KOC04}). Due to the location of the
lensed images behind the bulge of the lens, we assume that 100\% of
the matter is in the form of compact objects. This assumption is
motivated by the results of Kochanek (\cite{KOC04}), who demonstrate
that the most likely $\kappa_{\rm smooth}=0$ in \obj. The mass
function of microlenses has little effect on the simulations
(Wambsganss~\cite{WAM92}, Lewis \& Irwin~\cite{LEW95},~\cite{LEW96},
Congdon et al.~\cite{CON07}), so we assume identical masses in the
simulations. The mass of the microlenses sets the Einstein radius
$r_E$. For \obj, the Einstein radius projects onto the source plane
as

\begin{equation}
r_E = \sqrt{\frac{4G\avg{M}}{c^2}\frac{D_{os}D_{ls}}{D_{ol}}} = 9.84 \times 10^{16} \sqrt{\frac{\avg{M}}{0.3 M_{\sun}}} {\rm {cm}},
\end{equation}

\noindent where the $D$ are angular diameter distances, and the
indices $o$, $l$, $s$ refer to observer, lens and source, and
${\avg{M}}$ is the mass of microlenses. We create magnification
patterns with sidelength of 100 $r_E$ and pixel sizes of 0.01 $r_E$.
Tracks drawn in such a pattern would provide simulated microlensing
lightcurves for one pixel-size source. To simulate lightcurves for
other source sizes, we convolve the magnification pattern (after
conversion on a linear flux scale) by the intensity profile of the
source. As shown by Mortonson et al. (\cite{MOR05}), the exact shape
of the source intensity profile has little influence on the
lightcurve. It instead depends on the characteristic size of the
source. For simplicity, we assume a Gaussian profile. A uniform-disc
profile is also tested for comparison. We construct magnification maps
for 60 different source sizes having characteristic scales (FWHM for
the Gaussian and radius for the disc profile) in the range 2-1400
pixels (0.02 $r_E$ - 14 $r_E$).

\subsection{Comparison with the data}
\label{subsec:compa}

The patterns created in Sect.~\ref{subsec:pattern} are used to extract
simulated microlensing lightcurves to be compared to the data. Two
data sets are used, the OGLE data of \obj\, (Udalski et
al.~\cite{UDA06}), which provide well-sampled V-band lightcurves of
the continuum variations and our spectrophotometric lightcurves
presented in Sect.~\ref{sec:obs}. To study the microlensing signal, we
have to get rid of intrinsic variations. This is done by calculating
the difference lightcurves between images A \& D. Similarly, the
simulated microlensing signal is obtained by taking the differences
between the simulated tracks for images A \& D. The parameters that
characterise a simulated microlensing track are the starting
coordinates in the magnification patterns (${\bf {x_{0,A}}}$, ${\bf
  {x_{0,D}}}$), the track position angle ($\theta_A$, $\theta_D$), and
the track length. Because the extracted track is compared to a
lightcurve obtained over a given time range, the track length
corresponds to the transverse velocity of the source expressed in
$r_E$ per Julian day. As in Paper II, we assume that the velocity is
the same for the trajectories in A \& D and the track orientation
$\theta_A$= $\theta_D$. To account for the roughly orthogonal shears
in A \& D (Witt \& Mao \cite{WIT94}), we rotate the magnification map
of D by 90$^\circ$ prior to the track extraction. The difference of
the magnitude between track A and track D should, on average, be equal
to the one of the macro-model. Like other authors, we account for the
uncertainty on the amount of differential extinction between A \& D
and on the macro-model flux ratio by allowing for a magnitude offset
$m_0$ between the simulated lightcurves extracted for A and D. The
other parameters characterising the lightcurves are the convergence
and shear ($\kappa$,$\gamma$) fixed in
Sect.~\ref{subsec:pattern}. Physical quantities (source size and
velocity) are proportional to the mass of microlenses as
$\avg{M}^{1/2}$.

\begin{figure}
\begin{center}
\includegraphics[width=9.2cm]{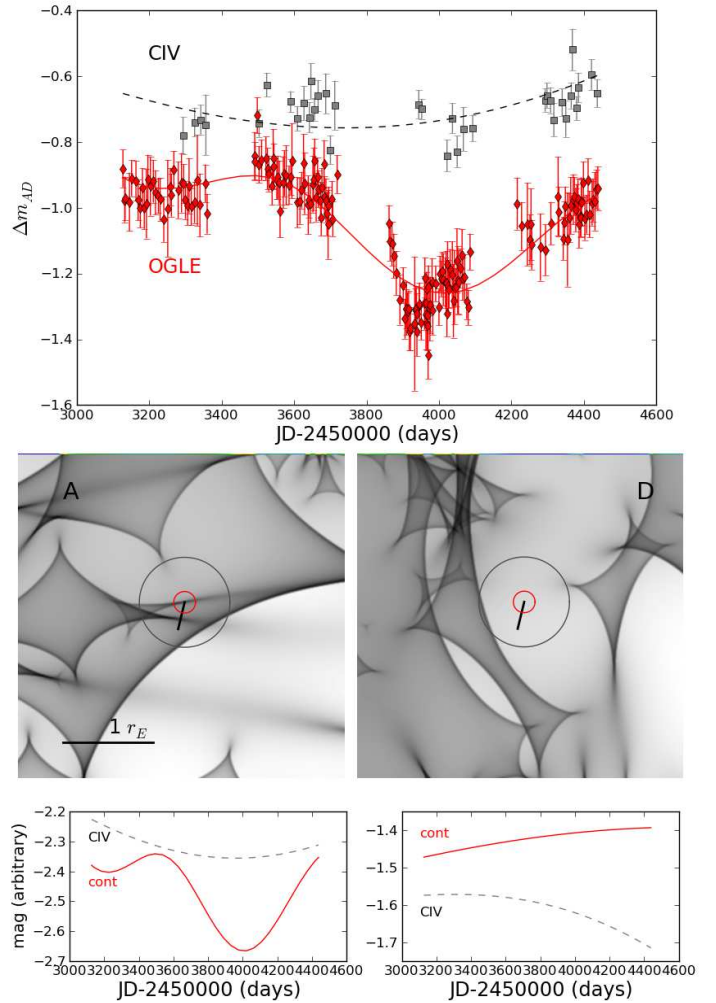}\\
\end{center}
\caption{ {\it {Top:}} Example of a good track fitting the differential lightcurve $\Delta m_{AD}$ observed for the \ion{C}{IV} broad line and for the continuum. {\it {Centre:}} Corresponding tracks (solid line) in the microlensing pattern of image A ({\it{left}}) and D ({\it{right}}).  The large circle represents the \ion{C}{IV} source size and the small one the continuum. {\it {Bottom:}} Track corresponding to the continuum emission (red solid line) and to the \ion{C}{IV} emitting region (dashed grey line) for image A (left) and D (right). }

\label{fig:bestfit}
\end{figure}

To build a representative ensemble of lightcurves that is compatible
with the data, we followed a four-step strategy similar to the one
described in Paper II, in Anguita et al. (\cite{ANG08}) and in
Kochanek (\cite{KOC04}): {\it {1)}} we pick a set of starting values
for the parameters defining the tracks in A \& D and vary them to
minimise the $\chi^2$ between the simulated lightcurve and the OGLE
lightcurve; {\it 2)} we repeat step 1 to get a representative ensemble
of 10000 tracks and good coverage of the parameter space; {\it 3)}
each track estimated in step 2 is computed for other source sizes, and
the agreement with the BLR lightcurve is quantified using a
$\chi^2$-type merit function; {\it 4)} the $\chi^2$ estimated in steps
(2) and (3) are summed up and used to calculate a likelihood
distribution for any track parameter. A summary of the technical
details regarding microlensing simulations is given in
Appendix~\ref{app:microfit}. An example of a good simulated lightcurve
fitting the data is shown in Fig.~\ref{fig:bestfit}. In the next
subsection, we explain how we derive probability distributions for the
quantities of interest.

\subsection{Probability distribution}
\label{subsec:Bayes}

The probability of a trajectory $j$, defined by the set of parameters
$p=(\kappa, \gamma, \avg{M}, R^c_{s}, R^l_{s}, V, m_0, {\bf
  {x_{0,A}}}, {\bf{x_{0,D}}}$) (where we drop the index $j$ for
legibility, $R^c_{s}$ and $R^l_{s},$ refer to the source size of the
continuum and of the line) given the data $D$ is, according to Bayes
theorem:
\begin{equation}
\label{equ:proba}
P(p_j|D) = \frac{L(D|p_j)P(p_{j})}{\sum^n_{j=1}{L(D|p_j)P(p_{j})}} = \frac{L(D|p_j)P(R^c_{s,j})P(V_j)}{\sum^n_{j=1}{L(D|p_j)P(R^c_{s,j})P(V_j)}}, \nonumber 
\end{equation}

\noindent where $n$ is the total number of trajectories in the
library, $L(D|p_j)$ is a likelihood estimator of the data $D$ given
the parameters $p_j$ and $P(R^c_{s,j})$ and $P(V_j)$ are the priors on
the continuum source size and velocity respectively. Only the prior on
$R^c_s$ and on $V$ appear in Eq.~\ref{equ:proba} because we use an
uniform prior on the other (nuisance) parameters. The prior
$P(R^c_{s,j})$ is required because we have a non uniform sampling of
$R^c_s$. The prior is related to the density of trajectories as a
function of $R^c_{s,j}$ given our source sampling in 60 different
bins. If $R^c_{s,j}$ falls in the bin $b$ of width $l_b$, then we have
$P(R_{s,j}) \propto l_b$. The velocity prior $P(V_j)$ is identical to
the one used in Anguita et al. (\cite{ANG08}). It is based on previous
studies by Gil-Merino et al. (\cite{GIL05}), which suggest that the
net transverse velocity is lower than 0.001 $r_E$/jd (at 90\%
confidence level). This corresponds to v$ < $ 660 km\,s$^{-1}$ in the
lens plane ($\sim$ 6000 km\,s$^{-1}$ in the source plane) for $\avg{M}
\sim 0.1 M_{\sun}$, in agreement with the velocity prior used by
Kochanek (\cite{KOC04}) and Poindexter \& Kochanek
(\cite{POI10a}).

As explained in Kochanek (\cite{KOC04}), the use of the standard
likelihood estimator $L(D|p_j)=exp(-\chi^2(p_j)/2)$ can lead to
misleading results because it is very sensitive to the accuracy of the
error bars and to change of the $\chi^2$ induced by possible
outliers. To circumvent this problem, we follow Kochanek
(\cite{KOC04}) and use the following likelihood estimator:

\begin{equation}
L(D|p_j) = \Gamma\left[\frac{n_{dof}-2}{2}, \frac{\chi^2(p_j)}{2}\right],
\end{equation}

\noindent where $\Gamma$ is the incomplete gamma function, and $\chi^2
=\chi^2_{OGLE}+\chi^2_{line}$ is the sum of the $\chi^2$ obtained when
fitting the OGLE data and emission line lightcurves (steps 2 \& 3
in Sect.~\ref{subsec:compa}).

\section{Inferences of the quasar structure}
\label{sec:results}

In this section, we present the results of our microlensing
simulations and discuss them in light of the phenomenological
study of the emission lines in A \& D. First, we discuss the domain of
validity of our simulations and the meaning of the BLR size estimated
with our method. Second, we derive the size of the continuum and of
the BLR for \ion{C}{IV} and \ion{C}{III]}. Third, we compare our
results to the radius(BLR)-luminosity relation derived from
reverberation mapping. Fourth, we derive the black hole mass of \obj\,
using the virial theorem and compare it to black hole mass derived
from the accretion disc size. Finally, we discuss the consequences of
our analysis for the BLR structure of \obj.

\subsection{Caveat}

In microlensing simulations, we represent the projected BLR with a
face-on disc having a Gaussian or a uniform intensity. The exact
intensity profile should not affect our estimate of the size
(Mortonson et al.~\cite{MOR05}), but the sizes retrieved are only
robust if the BLR actually has a morphology that is disc-like when
projected on the plane of the sky. This corresponds to the case of a
spherically symmetric BLR geometry or of an axi-symmetric geometry
with the axis roughly pointing towards us. Since the microlensing
signal remains nearly unchanged for sources with ellipticity
$e=1-q<0.3$ (Fluke \& Webster ~\cite{FLU99}, Congdon et
al.~\cite{CON07}), our simulations should also be valid for disc-like
sources inclined by up to $\phi \sim 45^\circ$ with respect to the
line of sight. This means that inclination may affect our size
estimates by a factor $\cos \phi > 1/\sqrt{2}$.  Although the exact
geometry of the BLR is unknown, there are several lines of evidence
that \obj\, is seen close to face-on.  Firstly, \obj\, is a Type 1 AGN
with a flat radio spectrum ($\alpha=-0.18$, Falco et
al.~\cite{FAL96}), indicating that the source is seen nearly
face-on. Second, the analysis of the microlens-induced continuum
flux variations by Poindexter \& Kochanek (\cite{POI10b}) show that
the inclination of the accretion disc is $<$45$^\circ$ at 68.3\%
confidence level. Third, the \ion{C}{IV} line shows only a small
blueshift that can be interpreted as evidence that the BLR is seen
face-on (Richards et al.~\cite{RIC02}).

If the BLR has a disc-like morphology, the projected emission should
take place in a ring-like{\footnote{In the case of emission coming
    from a spherical shell, the projected profile may also look
    ring-like, but emission should also arise from the central part of
    the disc.}}  region (with an inner radius $R_{in}$ and outer
radius $R_{out}$) rather than in a uniform disc as assumed here. The
effect of such a `hole' in the BLR emission has been studied by Fluke
\& Webster (\cite{FLU99}), who show that the microlensing lightcurve
would be significantly affected only for $R_{in}>0.5\,R_{out}$.  Since
the BLR is commonly assumed to have $R_{out} > 10\,R_{in}$ (e.g.
Murray \& Chiang~\cite{MUR95}, Collin \& Hur\'e~\cite{COL01}, Borguet
\& Hutsem\'ekers~\cite{BOR10}), we do not consider this geometrical
issue.

\subsection{Absolute and relative size of the continuum and the BLR}
\label{subsec:size}

\begin{table*}[tb]
\caption[]{Comparison of V-band continuum source size of \obj\, (Sect.~\ref{subsec:size}). The reference is presented in Col.~\#1, the source profile used in Col.~\#2, and the published size estimate in Col.~\#4 (where $R$= radius of a uniform disc, $R_V$= scale radius of a Shakura-Sunyaev accretion disc, $\sigma_V$ and FWHM are resp. the standard deviation and the full width at half maximum of a Gaussian). These quantities have been converted to half-light radius $R_{1/2}$ to allow comparison between profiles. Col.~\#6 \& \#7 provide the average mass of microlenses ($\avgg{M}$) used in the microlensing simulations and information regarding the velocity prior. }
\label{tab:compasizes}
\begin{center}
\begin{tabular}{lcllccl}
\hline 
\hline 
Reference   &   Profile & Estimator    &  Size (10$^{15}$cm) &  $R_{1/2}$ (10$^{15}$cm)& $\avgg{M}$ & Velocity prior \\
\hline
This paper  & Gauss                & Median & 17.6$^{+33.3}_{-12.5}$  (FWHM)   & 8.8$^{+16.6}_{-6.3}$ &  $\sqrt{\avg{M}/0.3 M_{\sun}}$ & Gil-Merino et al.\,(\cite{GIL05})  \\
This paper  & Uniform disc         & Median & 17.4$^{+34.6}_{-12.9}$  ($R$)   & 12.3$^{+24.5}_{-9.1}$ &  $\sqrt{\avg{M}/0.3 M_{\sun}}$  & Gil-Merino et al.\,(\cite{GIL05}) \\
Poindexter et al.\,(\cite{POI10b})       & Shakura-Sunyaev                &  Median &5.8$^{+3.8}_{-2.3} $  ($R_V$) & 14.1$^{9.2}_{-5.6}$& 0.3 $M_{\sun}$ & Poindexter et al.\,(\cite{POI10a}) \\
Anguita et al.\,(\cite{ANG08})       & Gauss  &  Most likely & 4.6$^{+3.4}_{-3.4}$  ($\sigma_V$) & 5.4$^{+4.0}_{4.0}$ & $\sqrt{\avg{M}/0.1 M_{\sun}}$ & Uniform  \\
Anguita et al.\,(\cite{ANG08})        & Gauss  &  Most likely & 1.3$^{+0.2}_{-0.7}$  ($\sigma_V$) & 1.5$^{+0.3}_{0.8}$ & $\sqrt{\avg{M}/0.1 M_{\sun}}$ & Gil-Merino et al.\,(\cite{GIL05})  \\
Paper II      & Gauss & Most likely & 40$^{+75}_{-35}$  (FWHM)       & 20$^{+37.5}_{-17.5}$ & $\sqrt{\avg{M}/0.1 M_{\sun}}$ & Uniform \\
Paper II       & Gauss & Most likely & 9.2$^{+6.9}_{-5.8}$   (FWHM)   & 4.6$^{+3.4}_{-2.9}$ & 0.1 $M_{\sun}$& Similar to  Kochanek (\cite{KOC04}) \\
Kochanek\,(\cite{KOC04})      & Shakura-Sunyaev    & Median  & 4.1$^{+6.7}_{-1.8}$  ($R_V$)    & 10$^{+16.3}_{-4.4}$   & $[0.1,1] M_{\sun}$ & Sect. 2.3 of Kochanek\,(\cite{KOC04}) \\
Kochanek\,(\cite{KOC04})        & Gauss        & Median  & 5.1$^{+4.7}_{-2.8}$  ($\sigma_V$)    & 5.9$^{+5.5}_{-3.2}$   & $[0.1,1] M_{\sun}$ & Sect. 2.3 of Kochanek\,(\cite{KOC04})\\

\hline
\end{tabular}
\end{center}
\end{table*}

\begin{figure}[tb]
\begin{center}
\includegraphics[width=8cm, bb = 0 170 550 600]{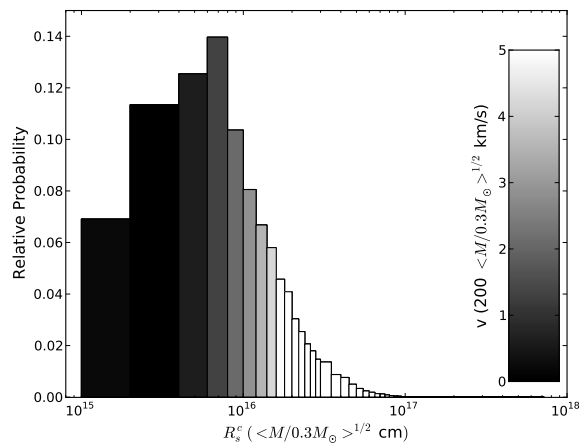} 
\caption{Probability distribution for the half-light radius of continuum emission in \obj, colour-coded as a function of the average (lens plane) velocity of the tracks in a given bin. }
\label{fig:probaRSsize}
\end{center}
\end{figure}

The V-band OGLE lightcurve is associated with emission arising mostly
from the quasar continuum.  The probability distribution of the
continuum source size $R^{c}_s$ derived from our microlensing study of
that lightcurve (Sect.~\ref{sec:simulation}) is displayed in
Fig.~\ref{fig:probaRSsize}. In this figure, we also show the average
velocity of the tracks in each source size bin on a colour-coded
scale. We see that the source size and the velocity are strongly
correlated as expected from the source size/velocity degeneracy
(e.g. Kochanek~\cite{KOC04}, Paper II). When we consider a uniform
prior on the velocity (not shown), we find that by imposing $m_0$, the
average difference of magnitude between A \& D lightcurves, and the
macro-model expectation to be equal, we exclude very high velocities
(v\,$> 0.005 r_E/$jd). For the velocity range allowed by the
Gil-Merino et al. (\cite{GIL05}) velocity prior, the same
distributions of $m_0$ are found regardless of what velocity bin is
considered. The median half-light radius of the continuum we derive
from our simulations is $R^{c}_s=8.8 \times 10^{15}\,$ cm ($ 2.5
\times 10^{15}\,$ cm $< R^{c}_s < 25.4 \times 10^{15}\,$ cm at 68.3\%
confidence level, $\avg{M}=0.3 M_{\sun}$) for a Gaussian profile, and
$R^{c}_s=12.3 \times 10^{15}\,$ cm ($ 3.2 \times 10^{15}\,$ cm $<
R^{c}_s < 36.8 \times 10^{15}\,$ cm at 68.3\% confidence level,
$\avg{M}=0.3 M_{\sun}$) for a uniform disc. Table~\ref{tab:compasizes}
compares our measurements with the results of Kochanek (\cite{KOC04},
KOC04), Anguita et al. (\cite{ANG08}), Paper II, and Poindexter \&
Kochanek (\cite{POI10b}). The half-light radius ($R_{1/2}$) allows the
comparison of sizes derived with various intensity profiles. We see
that our results are compatible with all other works.

\begin{figure}[tb]
\begin{center}
\begin{tabular}{c}
\includegraphics[width=8.cm, bb = 0 170 550 600]{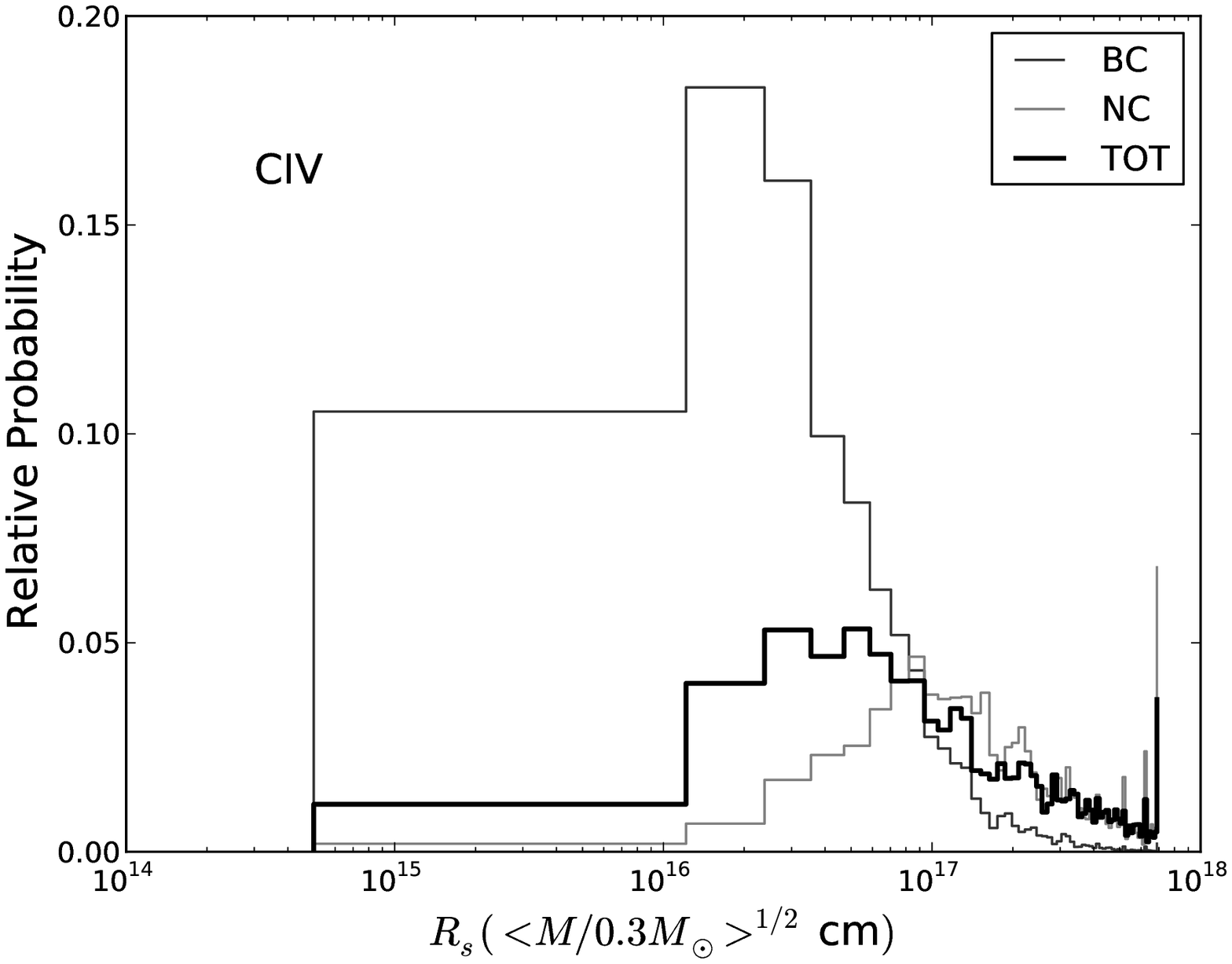} \\
\includegraphics[width=8.cm,  bb = 0 170 550 600]{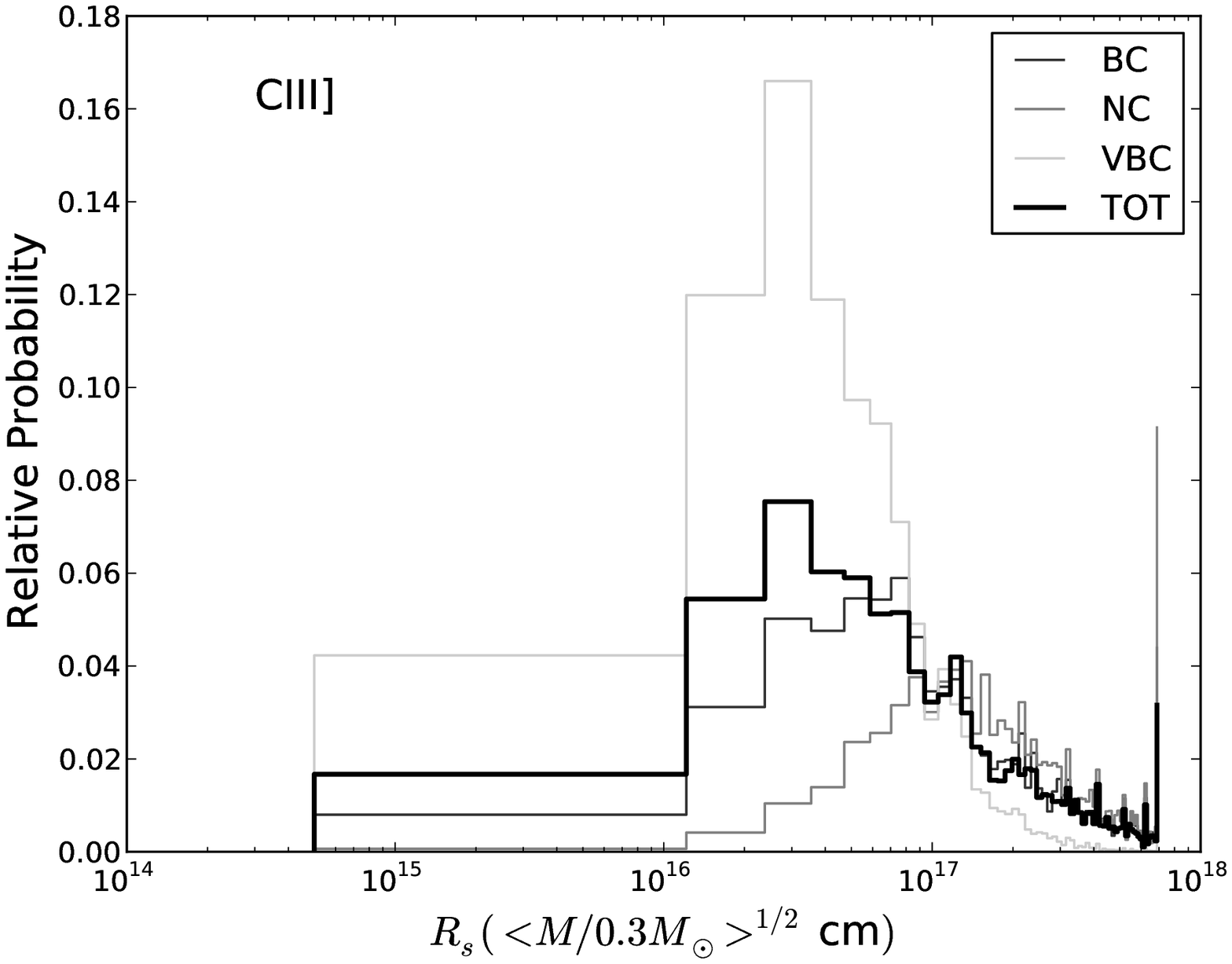} \\
\end{tabular}
\caption{Probability distribution for the size of the \ion{C}{IV} ({\it {top}}) and \ion{C}{III]} ({\it{bottom}}) broad emission lines. Measurements for the region emitting the \ion{C}{IV} \& \ion{C}{III]} full emission line profiles (thick black lines, TOT) and for their components NC, BC, and VBC (in order of increasing FWHM, cf. Fig.~\ref{fig:linedec}), are presented.}
\label{fig:probasize}
\end{center}
\end{figure}

The probability distributions of the half-light radius measured for
the \ion{C}{IV} and \ion{C}{III]} emission lines (using a Gaussian
source light profile) are displayed in
Figure~\ref{fig:probasize}. Reasonably good fits are obtained for all
the line components. However, the quality of the best fits tends to be
poorer for the broadest line component (typical reduced $\chi^2 \sim
2$). The poorer fits were obtained for the very broad component of the
\ion{C}{III]} line that had best reduced $\chi^2 \sim 4$. This
suggests that our error bars on the flux ratio are underestimated (by
a factor $\sim \sqrt 2$) when measured in the broadest line
components. This is expected since the flux in these components of the
emission lines are more prone to measurement errors in the MCD
decomposition. Our results are, however, robust in the sense that they
remain unchanged when the error bars are increased by this
factor. Median half-light radius and 68.3\% confidence intervals are
provided in Table~\ref{tab:size} for both Gaussian and uniform-disc
intensity profiles of the BLR. These sizes are independent of the
intensity profile used for the continuum emission. These results
confirm that the most compact region of the quasar is also the one
emitting the broadest components of the lines.

While the absolute source sizes derived above depend on the mass of
the microlenses, the relative size of the continuum and the BLR are
independent of the mass of microlenses. Table~\ref{tab:ratio} displays
the ratio between the emission line and the continuum half-light
radius size $R^{l}_s/R^{c}_s$ (assuming the same source profile for
the continuum and BLR) at 68.3\% confidence for the different
components of the BLR identified in Sect.~\ref{subsec:MCD}. We can see
that the region emitting the broadest components of the \ion{C}{IV}
and of the \ion{C}{III]} lines (i.e. \ion{C}{IV} BC and
\ion{C}{III]} VBC) are about 4 times larger than the region
emitting the V-band continuum.  In contrast, the narrowest component
of the emission lines arises from a region that is typically 25 times
larger than the continuum.

We notice that the radius of the extreme-UV continuum ionising
the carbon lines (typically $\lambda < 500$\,\AA) is expected to be
much smaller than the V-band continuum, with a dependence of the size
on wavelength as $R \propto \lambda^{\xi}$ (Paper II and
ref. therein). For a standard accretion disc model, $\xi=4/3$,
implying that the UV ionising continuum is more than 7 times smaller
than the OGLE-V band continuum.

Finally, we note that the results of Tables~\ref{tab:size} \&
~\ref{tab:ratio} are not sensitive to possible outliers in our BLR
lightcurves. We essentially retrieve the same results when the points
inconsistent with the OGLE lightcurve (i.e. those which are marked
with a 'x' in Fig.~\ref{fig:oglecompa}) are excluded. Similar results,
but affected by larger error bars, are also obtained if we do not
consider the 2007 data.

\begin{table}[htb]
\caption[]{Half-light radius of the line emitting region $ R^{l}_s$ (in $10^{15} \, \sqrt{\avg{M}/0.3\,M_{\sun}}$ cm) for a Gaussian (G) and a uniform disc (D) intensity profile. 
Measurements for the \ion{C}{IV} \& \ion{C}{III]} full emission line profiles (TOT) and for their individual components NC, BC and VBC (in order of increasing FWHM, cf Fig.~\ref{fig:linedec}), are presented.}
\label{tab:size}
\begin{center}
\begin{tabular}{lllll}
\hline 
\hline 
Line & $R^{l}_s$ (G)  & 68.3\% CI & $R^{l}_s$ (D) & 68.3\% CI \\
\hline
\ion{C}{IV} BC&42.9&[14.1,124.1] &56.6&[17.6,168.6] \\
\ion{C}{IV} NC&$>82.7$& - &$>103.5$&- \\
\ion{C}{IV} TOT&170.7&[47.7,457.6] &276.6&[72.0,676.0] \\
\ion{C}{III]} BC&150.1&[51.7,430.5] &209.6&[69.0,627.7] \\
\ion{C}{III]} NC&$>97.9$&- &$>112.2$&- \\
\ion{C}{III]} VBC&53.6&[21.2,126.9] &41.7&[17.4,103.0] \\
\ion{C}{III]} TOT&126.4&[36.9,399.5] &163.5&[43.8,570.6] \\
\hline
\end{tabular}
\end{center}
\end{table}

\begin{table}[htb]
\caption[]{Ratio of size $R^{l}_s/R^{c}_s$ of the BLR to continuum region for Gaussian (G) and uniform disc (D) source profiles. Measurements for the region emitting the \ion{C}{IV} \& \ion{C}{III]} full line profiles and for their components NC, BC and VBC (in order of increasing FWHM), as shown in Fig.~\ref{fig:linedec}, are presented. }

\label{tab:ratio}
\begin{center}
\begin{tabular}{lllll}
\hline 
\hline 
Line & $R^{l}_s/R^{c}_s$ (G) & 68.3\% CI & $R^{l}_s/R^{c}_s$ (D)& 68.3\% CI \\
\hline
\ion{C}{IV} BC    &3.9&[1.6,13.4]  &3.0&[1.1,12.0] \\
\ion{C}{IV} NC    &27.7&[9.0,89.3] &22.2&[7.1,73.8] \\
\ion{C}{IV} TOT   &18.9&[4.6,72.3] &17.5&[4.5,61.2] \\
\ion{C}{III]} BC  &16.9&[4.9,63.1] &15.2&[4.0,58.9] \\
\ion{C}{III]} NC  &29.4&[10.5,103.6] &25.6&[8.1,82.5] \\ 
\ion{C}{III]} VBC &5.0&[2.4,14.3]  &2.3&[0.9,7.9] \\
\ion{C}{III]} TOT &13.8&[0.7,53.8] &12.6&[3.4,49.4]  \\

\hline
\end{tabular}
\end{center}
\end{table}

\subsection{\ion{C}{IV} radius-luminosity relationship}
\label{subsec:relation}

\begin{figure}[tb]
\begin{center}
\includegraphics[width=9cm]{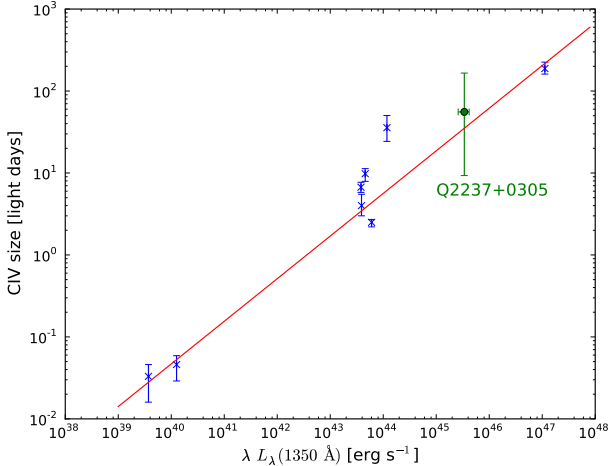} 
\caption{ $R_{\ion{C}{IV}}$-L diagram combining reverberation mapping measurement of $R_{\ion{C}{IV}}$ ('x' symbol, Peterson et al.~\cite{PET05}, Kaspi et al.~\cite{KAS07}) and our microlensing-based size for \obj. The solid line is the analytical fit to the reverberation mapping data from Kaspi et al. (\cite{KAS07}), which has a slope $\alpha = 0.52$.}
\label{fig:revsize}
\end{center}
\end{figure}

In this section, we compare our microlensing-based estimate of the
\ion{C}{IV} emission line (Sect.~\ref{subsec:size}) with the
radius(BLR)-luminosity relationship (hereafter $R_{BLR}-L$) derived
with the reverberation mapping technique (Sect.~\ref{subsec:RL}). The
$R_{BLR}-L$ relationship for \ion{C}{IV} has only been obtained for
six objects (Peterson et al.~\cite{PET05}, Kaspi et
al.~\cite{KAS07}). The luminosity used in this relationship is
commonly L$_\lambda$($1350\,\AA$). Unfortunately, the latter is out of
our spectral range. Instead we use
$\lambda$L$_\lambda$(1450$\,\AA$)=10$^{45.53}$\,erg/s (Assef et
al.~\cite{ASE10}), which has been shown to be equivalent to
L$_\lambda$($1350\,\AA$) (Fig. 4 of Vestergaard \&
Peterson~\cite{VES06}). This luminosity estimate, directly measured on
our spectra, is corrected for macro-magnification and line-of-sight
Galactic extinction. The macro-magnification is correct within a
factor 2 (Witt \& Mao~\cite{WIT94}). Because the lensed images are
located behind the bulge of the galaxy, it is reasonable to assume
that the absolute extinction due to the lens is $<$ 1.0 mag. The
finite slit width of 0.6$\arcsec$ and seeing $\sim$ 0.6$\arcsec$ may
lead to underestimating the total flux by a factor up to 2. This
leads to an uncertainty factor of 4 on
$\lambda\,$L$_\lambda$(1450$\,\AA$). The \ion{C}{IV} size has been
derived to be $R^{\ion{C}{IV}}_s= 1.7^{+2.9}_{-1.2} \times 10^{17}\,
\sqrt{\avg{M}/{0.3 M_{\sun}}}\,{\rm {cm}}$ =
$65.6^{+110}_{-46}\,\sqrt{\avg{M}/{0.3 M_{\sun}}}$\,light-days
(Table~\ref{tab:size}). Figure~\ref{fig:revsize} shows the $R_{\rm
  BLR}$-L diagram that includes our measurement for \obj\, and the
analytical relation derived by Kaspi et al. (\cite{KAS07}) for their
so-called 'FITEXY' fit. Although the reverberation-mapping size may
differ from the half-light radius by a factor up to a few depending on
the geometry and density of the BLR gas, we find remarkably good
agreement with the R-L$^{\alpha}$ relationship.

\subsection{Black hole mass estimates}
\label{subsec:MBH}

The BLR size we derive from microlensing can be used to estimate a
black hole mass using the virial theorem:

\begin{equation}
\label{equ:virial}
M_{BH}= \frac{f}{G} R_{BLR} \,FWHM^2_{BLR},
\end{equation}

\noindent where $f$=1 is a geometric factor, which encodes assumptions
regarding the BLR geometry (e.g. Vestergaard \& Peterson~\cite{VES06},
Collin et al.~\cite{COL06}), calibrated from the results of
Onken{\footnote{Onken et al. derive $f=5.5$ but using $\sigma^2$
    instead of FWHM$^2$ in the virial relation. The corresponding
    value of $f$ when using the FWHM is $f=1$.}}  et
al. (\cite{ONK04}). Using FWHM$_{\ion{C}{IV}} = 3960 \pm
180$\,km\,s$^{-1}$ (Assef et al. 2010), we derive $M_{BH}=
2.0^{+3.4}_{-1.5} \times 10^8\, M_{\sun}$. Consistent measurements, but
generally affected by larger error bars, are derived using the other
broad components of the emission lines. Hereafter we refer to this
black hole mass estimate as $M^{BLR}_{BH}$. As we argued earlier,
inclination with respect to the line of sight may reduce $R_{BLR}$ by
$\cos \phi > 1/\sqrt 2$.

Our estimate of $M^{BLR}_{BH}$ may be uncertain by a factor of a few
because of the variation of $f$ from object to object. A value of $f$ higher
than our fiducial value $f=1.0$ may be expected for axisymmetric
geometry of the BLR seen nearly face-on. The study of several
plausible theoretical models of the BLR allowed Collin et
al. (\cite{COL06}) to show that inclination of axisymmetric BLR
models may noticeably increase $f$. Observationally,
Decarli{\footnote{Decarli et al. use a different definition of $f$:
    $f_{\rm{this paper}}$=$f^2_{\rm{Decarli}}$.}} et
al. (\cite{DEC08}) find $f$ in the range [0.5,10], with higher values
($f>4$) only for objects with FWHM $<$ 4000\,km\,s$^{-1}$.

The use of the virial theorem to derive $M_{BH}$ based on the
\ion{C}{IV} line is still debated. For this line, non gravitational
effects, such as obscuration and radiation pressure, are probably
affecting the \ion{C}{IV} profile (Baskin \& Laor~\cite{BAS05},
Marziani et al.~\cite{MAR06}). Obscuration, if present, should lead to
underestimating FWHM$(\ion{C}{IV})$ and therefore to overestimating
the $M^{BLR}_{BH}$. The effect of radiation pressure might be
subtler. While Marconi et al. (\cite{MAR08}) argue that radiation
pressure implies an underestimate of $M_{BH}$, recent calculations of
Netzer \& Marziani (\cite{NET10}) show that orbits of BLR clouds are
affected by radiation pressure in such a way that the product
$R_{BLR}\,$FWHM$^2$ (and therefore $M_{BH}$) will vary by at most 25\%
owing to radiation pressure. Our finding of roughly similar $M_{BH}$ for
different components of the line, emitted in regions differently
affected by radiation pressure, further supports the idea that
$M_{BH}$ in \obj\ is not strongly biased by nongravitational effects.

A black hole mass estimate can also be derived from the size of the
accretion disc. In the accretion disc theory, the radius $R_\lambda$
which the disc temperature equals the photon energy, $kT =
hc/\lambda_{\rm rest}$ is given by the expression
(Poindexter \& Kochanek~\cite{POI10b})

\begin{equation}
\label{equ:Rdisc}
R_\lambda = \left(\frac{45\,G\,{\lambda}^4_{\rm rest}\,M_{BH}\,\dot{M}}{16\,\pi^6\,h\,c^2} \right)^{1/3},
\end{equation}

\noindent where $M_{BH}$ is the black hole mass and $\dot M$ is the accretion
rate. Rewriting this expression for the black hole mass, we get

\begin{equation}
\label{equ:disc}
M^{\rm disc}_{BH} = 0.68 \times \left(\frac{R_{1/2}}{10^{15}cm}\right)^{3/2} \, \left(\frac{\lambda_{\rm rest}}{0.2\,\mu m}\right)^{-2}\, \eta^{1/2}_{0.1}\left(\frac{L}{L_E}\right)^{-1/2}\, 10^8\,M_{\sun},
\end{equation}
\noindent where $0.1\,\eta_{0.1} = \eta = L/(\dot{M}c^2)$ is the
radiative efficiency of the accretion disc, $R_{1/2} = 2.44\,
R_{\lambda}$ the half-light radius of a Shakura-Sunyaev accretion
disc model, and $L/L_E$ the luminosity in units of Eddington
luminosity. Using the accretion disc half-light radius $R^{c}_s=8.8
\times 10^{15}\,$ cm (Sect.~\ref{subsec:size}), we get $M^{disc}_{BH} =
17.7 \times 10^8\,M_{\sun}$ (2.7$\times 10^8\,M_{\sun} < M^{disc}_{BH}
<$8.7$\times 10^9\,M_{\sun}$ at 68.3\% confidence) if we assume that $L=L_E$
and $\eta = $ 0.1. Unfortunately, $M^{\rm disc}_{BH}$ is also prone
to systematic errors and depends strongly on the exact temperature
profile of the disc. The black hole mass derived with this method
assumes a temperature profile $T \propto R^{-\xi'}$ with $\xi'=3/4$
(Shakura \& Sunyaev ~\cite{SHA73}). A different value of $\xi'$ in the
range [0.5,1] may lead to higher/lower $M^{disc}_{BH}$ by a factor
as large as $\sim 5$ (e.g. Poindexter \& Kochanek~\cite{POI10b}).  The
$M^{disc}_{BH}$ is thus more uncertain and less accurate than
$M^{BLR}_{BH}$.

By comparing the isotropic luminosity $L=4.0\, \times 10^{46}$
erg/s of \obj\,(Agol et al.~\cite{AGO09}) to the Eddington luminosity
$L_E = 1.38\,\times 10^{38}\,M_{BH}/M_{\sun}$ erg/s, we derive a
black hole mass $M^E_{BH} > 2.9 \, \times 10^{8} M_{\sun}$,
consistent with $M^{BLR}_{BH}$ and $M^{disc}_{BH}$.

\subsection{The structure of the BLR}

Based on the MCD decomposition of the spectra, we have shown that the
\ion{C}{IV} and \ion{C}{III]} lines can be decomposed into two
(resp. three) components of increasing width
(Fig.~\ref{fig:linedec}). The possibility that these line components
correspond to physically distinct regions in the BLR has been
discussed by several authors (e.g. Brotherton et al.~\cite{BRO94b},
Marziani et al.~\cite{MAR10}) but remains debated.  The variable
amount of microlensing in these components allows us to shed new light
on this scenario.

The differential microlensing lightcurves observed for the Gaussian
components of the emission line (Fig.~\ref{fig:MCD}) are essentially
flat, even when a substantial microlensing signal is observed in the
continuum. They also show an increasing microlensing effect in the
broadest line components. Our simulations show that it is possible to
reproduce both the continuum and BLR lightcurves by only changing the
source size. We clearly find that the most compact regions have the
broadest emission. This excludes BLR models for which the width of the
emission line is proportional to the size of the emitting region. Our
results are compatible with a variation in the BLR size with
FWHM$^{-2}$ as expected if the emitting regions are virialised. This
arises in various geometries, i.e., for a spherically isotropic BLR, a
Keplerian disc, or a disc+wind model similar to one proposed by Murray
\& Chiang (\cite{MUR95}, \cite{MUR97}).

The other techniques we used to analyse the spectra provide additional
evidence that the BLR of \ion{C}{IV} and \ion{C}{III]} is composed of
several regions.  The ``narrow-band'' technique
(Sect.~\ref{subsec:bandemission}) reveals that $\Delta m_{AD}$ is
closer to the macro model value (i.e., $\Delta m_{AD} \sim -0.1 \pm
0.1$) in the core of the emission lines than in the wings. But even in
the central 200\,km\,s$^{-1}$ part of the line, we find $\Delta
m_{AD}\sim$-0.6 mag, indicating that there is microlensed flux in the
core of the line. The MmD technique (Sect.~\ref{subsec:FMFMmu})
provides more clues to the origin of this flux. The shape of
$F_{M\mu}$ for the carbon lines smoothly increases from the base to
the line centre, and is compatible with a single-peak profile. The
lack of ``accretion disc-like'' profile with a flatter or depleted
core (see e.g. Bon et al. ~\cite{BON09}) does not exclude a disc-like
geometry of the BLR because disc emission can also produce
single-peaked profiles (Robinson~\cite{ROB95}, Murray \&
Chiang~\cite{MUR97}, Down \& al.~\cite{DOW10}). The MmD supports the
use of a Gaussian to empirically decompose the emission lines with the
MCD technique and a composite model of the BLR formed by at least 2
different regions.

Clues to the physical conditions occurring in these regions should
come from the flux ratios measured between lines. A detailed analysis,
however, needs very high signal-to-noise measurements (based e.g. on
an average spectrum of \obj\,), proper account of Galactic reddening,
and detailed photo-ionisation models. This is beyond the scope of this
paper and will be devoted to a future publication. We do, however,
mention several features already visible in our spectrum that should
provide interesting constraints on photo-ionisation models. First,
there is no broad \ion{Si}{III]} emission. The ratio
\ion{Si}{III]}/\ion{C}{III]} is very sensitive to the electronic
density (Keenan et al. ~\cite{KEE87}, Aoki \& Yoshida
~\cite{AOK99}). The absence of \ion{Si}{III]} in the broad component
of the emission indicates that the electronic density $n_e$ in the
region emitting the broad component of \ion{C}{III]} is lesser than in
the region emitting the narrow portion of the lines. Second, there is
no clear narrow \ion{Fe}{II+III} emission, suggesting that the latter
arises from the inner part of the BLR or from a very compact
region. This contrasts with microlensing observed in the lensed quasar
RXJ 1131-1231 (Sluse et al. 2007) where narrow and broad emission were
found for the `optical' \ion{Fe}{II+III} emission ($\lambda\,\in
[3000, 5000]\,\AA$), since the UV \ion{Fe}{II} is poorly covered by the
spectra in that system. Third, the ratio \ion{C}{III]}/\ion{C}{IV} is
found to be nearly the same in the broad and narrow components of the
line (Fig.~\ref{fig:velocity}). This advocates for a similar
ionisation parameter U in these two regions (Mushotzky \& Ferland
~\cite{MUT84}), which is {\it {a priori}} surprising since U should
vary with the distance. In their decomposition of the emission lines
based on the principal component analysis technique, Brotherton et
al. (\cite{BRO94b}) faced a similar problem and found that
\ion{C}{III]}/\ion{C}{IV} measured for the narrow part of the line
(their so-called intermediate line region) was too large by an order
of magnitude compared to photo-ionisation models. We speculate that
metallicity and optically thick/thin nature of the BLR gas are likely
to play important roles in reproducing these ratios (Snedden \&
Gaskell~\cite{SNE99}).

Small differences between the \ion{C}{IV} and \ion{C}{III]} line
profiles are visible in the total line profile
(Fig.~\ref{fig:velocity}). The \ion{C}{III]} profile peaks nearly at
the systemic redshift and is rather symmetric, while the \ion{C}{IV}
line shows a low blueshift and an excess of emission in the blue
(i.e. blue asymmetry), in agreement with previous studies (Brotherton
et al.~\cite{BRO94a}). Figure~\ref{fig:velocity} shows that these
differences are visible in the narrow component of the line. On the
other hand, the very broad component of these lines is slightly
asymmetric for both lines, showing again a blue excess but a more
greater one for \ion{C}{IV}. Because the intrinsic absorber is
superimposed on the \ion{C}{IV} profile, it is hard to assess whether
the broadest component of the profile is in fact blueshifted as is the
narrow one. Finally, we mention that the core of the carbon lines
varies less than the base of the line, in agreement with the finding
of Wilhite (\cite{WIL06}) based on a large sample of SDSS
quasars. This finding is consistent with the above arguments showing
that the broadest part of the line arises from a region closer to the
continuum.

\section{Conclusions}
\label{sec:conclusions}

For the first time, we have derived the size of the broad line region
in \obj\, based on spectrophotometric monitoring data and consisting
of 39 different epochs obtained between Oct 2004 and Dec 2007. To
reach this goal, we measured differential lightcurves between images A
\& D for the \ion{C}{IV} and \ion{C}{III]} broad emission lines and
compared them to microlensing simulations. This led to determining the
half-light radius of the \ion{C}{IV} emitting region: $R_{\ion{C}{IV}}
\sim $ 66 lt-days (19 lt-days $< R_{\ion{C}{IV}} <$ 176 lt-days at
68.3\% confidence), in very good agreement with the $R_{BLR}-L$
relation derived from reverberation mapping (Kaspi et
al.~\cite{KAS07}). The size we derived for \ion{C}{III]} is
$R_{\ion{C}{III]}} \sim 49$ lt-days (14 lt-days $< R_{\ion{C}{III]}}
<$ 154 lt-days at 68.3\% confidence), compatible with the size
estimated for the region emitting \ion{C}{IV}.

Thanks to the variable amount of microlensing observed within a given
emission line, we can also derive information on the structure of the
broad line region. Differential lightcurves obtained for various
velocity slices of the \ion{C}{IV} and \ion{C}{III]} lines show that
the wings of the lines are more microlensed than the core, indicating
that the former arise in a more compact region. This finding is
confirmed by two other techniques. The first one demonstrates that a
broad and single-peaked fraction of the emission line is microlensed,
while a narrower fraction is unaffected by microlensing.  This
technique also suggests a slightly different structure for the
\ion{C}{IV} and \ion{C}{III]} emission regions, with the narrow
\ion{C}{IV} emission blueshifted with respect to the systemic
redshift. The second technique assumes that the emission lines can be
decomposed into a sum of Gaussian components. The \ion{C}{IV} is
separated into a narrow (FWHM $\sim$ 2600\,km\,s$^{-1}$) and a broad
(FWHM $\sim$ 6300\,km\,s$^{-1}$) component. In order to reproduce the
small differences between the \ion{C}{IV} and \ion{C}{III]} lines, we
need three Gaussian profiles (FWHM $\sim$ 1550, 3400,
8550\,km\,s$^{-1}$) to reproduce the latter. The lightcurves derived
for these components are essentially flat and show that microlensing
is more important when the FWHM increases. Although the individual
Gaussian line components do not necessarily isolate individual
emission regions, we find that these lightcurves are compatible with
the continuum lightcurves, provided only the size of the emission
region is modified. This allows us to derive a half-light radius for
the regions emitting these components, as well as for their size
relative to the continuum. The radii are consistent with the virial
hypothesis and a radius that varies as FWHM$^{-2}$. Using the virial
theorem, we derived a black hole mass $M_{BH} \sim $2.0$\times 10^8
M_{\sun}$ (0.5$\times 10^8 M_{\sun} < M_{BH} < $ 5.4$\times 10^8
M_{\sun}$ at 68.3\% confidence).

Our analysis supports the findings by other authors (Brotherton et
al.~\cite{BRO94a}, Marziani et al.~\cite{MAR10}) that the regions
emitting the \ion{C}{IV} and \ion{C}{III]} lines are composed of at
least two spatially distinct components, one emitting the narrow core
of the line and another, more compact, emitting a broadest
component. The broad (resp. narrow) component of the \ion{C}{IV} and
\ion{C}{III]} lines do not have exactly the same profiles. The flux
ratio \ion{C}{III]}/\ion{C}{IV} is very similar when measured in the
broad and in the narrow components of the lines. This suggests that
the ionisation parameter U is nearly the same in the two regions, a
surprising result since the narrow components are found to arise in
regions that are several times larger than the broad components. Other lines
observed in our spectra seem to arise from a least two components:
\ion{Mg}{II}\,$\lambda 2798$, \ion{Al}{II}\,$\lambda 1671$,
\ion{He}{II}\,$\lambda 1640$, \ion{Al}{III}\,$\lambda 1857$. The
situation is different for \ion{Si}{III]}\,$\lambda 1892$, which does
not show emission from a broad component, which is indicative of a smaller
electronic density $n_e$ in the region emitting the broadest part of
the emission lines. On the other hand, we detect broad microlensed
\ion{Fe}{II+III} but no ``extended'' emission.  This suggests that
Fe$_{UV}$ is produced in the inner part of the BLR or in a very
compact region.  Obtaining spectrophotometric monitoring data in the
near-infrared where Balmer lines are detectable would be very useful
for constraining photoionisation models and comparing the microlensing
signal in Balmer (e.g. \ion{H}{$\beta$}), high ionisation
(e.g. \ion{C}{IV}), and Fe$_{opt}$ lines.

We demonstrated that the spectrophotometric monitoring of microlensing
in a lensed quasar is a powerful technique for probing the inner
regions of quasars, measuring the size of the broad line region, and
infering its structure. More work is still needed to take full
advantage of the method, but our results are very promising. Several
improvements are possible to increase the accuracy of our size
measurements and better characterise the BLR structure. First,
microlensing simulations reproducing the signal in more than 2 lensed
images should allow one to narrow the final probability distribution
on the source size. Second, the implementation of spectral fitting
using Markov-Chain Monte-Carlo should allow a more appropriate
estimate of the error bars and more advanced modelling of the
individual spectral components. Third, a fully coherent scheme should
be developed to consistently model the emission line shape and the
corresponding source intensity profile used in the simulation. These
improvement will be the subject of a future work.

\begin{acknowledgements}
  We thank B.~Borguet for useful discussions, and the referee,
  T.~Boroson, for valuable suggestions. DS acknowledges support from
  the Humboldt Foundation. This project is partially supported by the
  Swiss National Science Foundation (SNSF).
\end{acknowledgements}

\newpage
\appendix
\section{Averaged emission line profile}
\label{sec:emiline}

We explain here how we construct an averaged emission line profile and
how we use it to choose a reasonable multi-Gaussian decomposition of
the broad emission lines.  Following the prescriptions of the
multi-component decomposition described in Sec.~\ref{subsec:MCD}, we
are able to subtract the quasar pseudo-continuum emission (including
the \ion{Fe}{II+III} emission) from the quasar spectra. This procedure
leads to a spectrum containing only the broad emission lines. Then, we
normalize their flux and compare the line profiles averaged over
periods P1, P2, and P3. This procedure reveals that the profile of the
\ion{C}{III]} emission line basically remains unchanged over the whole
monitoring period. In contrast, there is evidence of a small change in
the \ion{C}{IV} line profile between 2005 and 2007. The high signal-
to-noise emission line spectra of \ion{C}{III]} and \ion{C}{IV} are
then fitted with a sum of 3 Gaussian-line profiles. For \ion{C}{III]},
we find that the best fit is obtained with a narrow emission component
with FWHM $=$ 1545\,km\,s$^{-1}$, an intermediate emission component
FWHM $=$ 3400\,km\,s$^{-1}$, and a broad component with
FWHM $=$ 8548\,km\,s$^{-1}$.  For \ion{C}{IV}, the emission line is well
fitted with only two emission components with FWHM $\sim$
2570\,km\,s$^{-1}$ and FWHM $\sim$ 6150\,km\,s$^{-1}$. The absorption
component is also fitted with an FWHM $\sim $ 1250\,km\,s$^{-1}$ for P1
and P2 and FWHM$ \sim $ 900\,km\,s$^{-1}$ for P3. The difference in
width of the absorption comes from the increased resolution of the
spectra obtained during period P3. Higher resolution spectra (Hintzen
et al. \cite{HIN90}, Yee \& De Robertis \cite{YEE91}) show that the
\ion{C}{IV} absorption system is composed of at least two different
absorption clouds associated with the quasar. Narrow absorption
systems are also present in front of \ion{C}{III]} emission (Motta et
al.~\cite{MOT04}), but they do not appear as significant features in
our spectra and are therefore not part of our modelling.

If we now compare the decomposition of the line profile in images A \&
D, we find that the width of the individual Gaussian profiles are
similar but not their relative intensities, as already suggested by
the analysis in Paper I. During our fitting procedure, we assume that
the FWHM of the Gaussian components are identical for A \& D, except
for \ion{C}{IV}(BC) and \ion{C}{IV}(AC). The widths are also fixed to
the fiducial values derived here.

\section{Microlensing fit}
\label{app:microfit}

We provide here the technical details regarding the comparison of the
simulated tracks with data. The following source sizes are considered:
2, 4, 8, 12, 16, 20, 24, 28, 32, 36, 40, 44, 48, 52, 56, 60, 70, 80,
90, 100, 110, 120, 130, 140, 150, 160, 170, 180, 190, 200, 240, 280,
320, 360, 400, 440, 480, 520, 560, 600, 640, 680, 720, 760, 800, 840,
880, 920, 960, 1000, 1040, 1080, 1120, 1160, 1200, 1240, 1280, 1320,
1360, and 1400 pixels. For each of the 60 source sizes $R_s$, we
simulate 10000 pairs of lightcurves for both images A \& D by tracing
source trajectories across the magnification patterns. Bilinear
interpolation is used for the track extraction. Each track is
characterised by the set of parameters described in
Sect.~\ref{subsec:compa}: $p=(R_s, V, m_0, \theta, {\bf{x_{0,A}}},
{\bf{x_{0,D}}}, \avg{M})$. Each simulated difference lightcurve
$\Delta m'_k(p)$ is then compared to the data $D = \Delta m_k$ by
measuring the goodness of fit with a $\chi^2$ statistic,
\begin{equation}
\chi^2(p) = \sum_{k=1}^{n_{\rm obs}} {\left(\frac{\Delta m_k - \Delta m_k'(p)}{\sigma_k}\right)^2},
\label{eq:chi2}
\end{equation}
where $\sigma_k$ are the uncertainties of the data and $n_{\rm obs}$
is the number of data points. For OGLE data, we use those obtained
between modified julian day (i.e. jd-2450000) 3100 and 4500, and we
bin the data points obtained the same night, which leads to $n_{\rm
  obs} = 181$. The error $\sigma_k$ is the quadratic sum of the
photometric error $\sigma_{\rm OGLE}$ provided by OGLE, of the
standard deviation $\sigma_{\rm bin}$ between binned points, and of
the systematic underestimation of the error $\sigma_{sys}$ of the OGLE
data. The error $\sigma_{sys}$ is estimated using a polynomial fitting
of the difference lightcurves (see Paper II) and is estimated to be
$\sigma_{sys} = $ 0.05 mag. The data points corresponding to the BLR
lightcurve have been described in Sect.~\ref{sec:flux}.

The brute force approach, which consists in randomly picking track
parameters and estimating a $\chi^2$ is very inefficient in getting
even a few tracks reproducing the data unless a large amount of
computing time is available. To overcome this problem, we follow the
procedure explained in Paper II and, for each set of random parameters
we have generated, we optimise the 6 parameters ($m_0, {\bf{x_A}},
{\bf{x_D}}, \theta$) with a $\chi^2$ based minimisation algorithm
using a Levenberg-Marquardt least-square routine. Contrary to what was
done in Paper II, we do not optimise the track velocity to avoid
possible bias on the velocity distribution induced by the minimisation
routine. In addition, we windowed the magnification map in such a way
that the starting point of a track is always more then one track length
away from the border of the map. Our final library of tracks fitting
OGLE data contains 6$\times 10^5$ tracks (10000 track per source size)
with a $\chi^2/n_{obs} < 8.5$.

The only parameter that is varied to compare the tracks of our
library to the BLR lightcurves is the source size $R_s$. As explained
in Sect.~\ref{subsec:compa}, we extract, for each track of our library
(step 2), the same track in the 59 other magnification maps
corresponding to the 59 source sizes. We then search for the best
source size using a minimisation routine and calculate tracks for
arbitrary source sizes (in the range 0.01-14 $r_E$) by means of a
quadratic spline interpolation between different source sizes.


\begin{thebibliography}{}

\bibitem[2002]{ABA02}       Abajas, C., Mediavilla, E., Mu\~noz, J.~A. et al. 2002, ApJ, 576, 640
\bibitem[2007]{ABA07}   Abajas, C., Mediavilla, E., Mu{\~n}oz, J.~A., G\'omez-\'Alvarez, P., Gil-Merino, R., 2007 ApJ, 658, 748
\bibitem[2009]{AGO09} Agol, E., Gogarten, S.~M., Gorjian, V., Kimball, A. 2009, ApJ 697, 1010
\bibitem[2008]{ANG08} Anguita, T., Schmidt, R.~W., Turner, E.~L. et al. 2008, A\&A, 480, 327
\bibitem[1999]{AOK99} Aoki, K., Yoshida, M. 1999, ASPC 162, 385
\bibitem[2010]{ASE10} Assef, R.~J., Denney, K.~D., Kochanek, C.~S. et al. 2010, ApJ, submitted (arXiV:1009.1145v1)
\bibitem[2004]{BAC04} Bachev, R., Marziani, P., Sulentic, J.~W. et al. 2004, ApJ, 617, 171
\bibitem[2004]{BAS04} Baskin, A., Laor, A., 2004, MNRAS 350, L31
\bibitem[2005]{BAS05} Baskin, A., Laor, A., 2005, MNRAS 356, 1029
\bibitem[2006]{BEN06} Bentz, M.~C., Peterson, B.~M., Pogge, R.~W., Vestergaard, M., Onken, C.~A. 2006, ApJ, 644, 133
\bibitem[2009]{BEN09b}	Bentz, M.~C., Walsh, J.~L., Barth, A.~J. 2009, 	ApJ, 705, 199
\bibitem[2010]{BEN10}   Bentz, M.~C., Horne, K., Barth, A.~J. et al., 2010, ApJ, 720, L46
\bibitem[2009]{BON09} Bon, E., Popovi{\'c}, L.~{\v C}., Gavrilovi{\'c}, N., Mura, G.~L., Mediavilla, E. 2009 MNRAS, 400, 924
\bibitem[2010]{BOR10} Borguet, B., Hutsem\'ekers, D. 2010, A\&A, 515, 22 
\bibitem[1992]{BOR92} Boroson T.~A., Green, R.~F. 1992, ApJS, 80, 109
\bibitem[1994a]{BRO94a} Brotherton, M.~S., Wills, B.~J., Steidel, C.~C., Sargent, W.~L.~W. 1994a ApJ, 423, 131
\bibitem[1994b]{BRO94b} Brotherton, M.~S., Wills, B.~J., Francis, P.~J., Steidel, C.~C., 1994b ApJ, 430, 495
\bibitem[1989]{CAR89} Cardelli, J.~A., Clayton, G.~C., Mathis, J.~S., 1989, ApJ, 345, 245
\bibitem[1989]{CHE89} Chen, K., Halpern, J.~P. 1989, 	ApJ, 344, 115 
\bibitem[1991]{CLA91} Clavel J., Reichert, G.~A., Alloin, D. et al. 1991, ApJ, 366, 64 
\bibitem[2001]{COL01} Collin, S., Hur\'e, J.-M. 2001, A\&A, 372, 50
\bibitem[2006]{COL06}  Collin, S., Kawaguchi, T., Peterson, B.~M., Vestergaard, M. 2006 A\&A, 456, 75
\bibitem[2000]{COU00}      Courbin, F., Magain, P., Kirkove, M., Sohy, S. 2000, ApJ, 529, 1136
\bibitem[2007]{CON07}   Congdon, A.~B., Keeton, C.~R., Osmer, S.~J. 2007, MNRAS, 376, 263 
\bibitem[1990]{COR90}   Corbin, M.~R. 1990, ApJ, 347, 346
\bibitem[1995]{COR95}   Corbin, M.~R. 1995, ApJ, 447, 496
\bibitem[1998]{COR98}   Corbett, E., Robinson, A., Axon, D.~J., Young, S., Hough, J.~H. 1998, MNRAS, 296, 721
\bibitem[2008]{DEC08} Decarli, R., Labita, M., Treves, A., Falomo, R. 2008, MNRAS, 387, 1237
\bibitem[2009]{DEN09} Denney, K.~D., Peterson, B.~M., Pogge, R. W. et al. 2009, ApJ, 704, L80
\bibitem[2003]{DIE03}     Dietrich, M., Hamann, F., Appenzeller, I., et al. 2003, ApJ, 596, 817
\bibitem[2010]{DOW10}  Down, E. J., Rawlings, S., Sivia, D. S., Baker, J. C., 2010, MNRAS, 401, 633
\bibitem[2007]{EIG07} Eigenbrod, A., Courbin, F., Sluse, D., et al. 2007, A\&A, 480, 647 (Paper I)
\bibitem[2008]{EIG08} Eigenbrod, A., Courbin, F., Meylan, G., et al. 2008, A\&A, 490, 933 (Paper II)
\bibitem[2000]{ELV00} Elvis, M. 2000, ApJ, 545, 63
\bibitem[1996]{FAL96}   Falco, E.~E., Lehar, J., Perley, R.~A., et al. 1996, AJ, 112, 897 
\bibitem[1999]{FAL99}   Falco, E.~E., Impey, C.~D., Kochanek, C.~S. et al. 1999, ApJ, 523, 617
\bibitem[2010]{FIN10} Fine, S., Croom, S.~M., Bland-Hawthorn, J. et al. 2010, MNRAS, 409, 591
\bibitem[1992]{FRA92}        Francis, P.~J., Hewett, P.~C., Foltz, C.~B., Chaffee, F.~H. 1992, ApJ, 398, 476
\bibitem[1999]{FLU99}    Fluke, C.~J., Webster, R.~L. 1999, MNRAS, 302, 68
\bibitem[1982]{GAS82}        Gaskell, C.~M. ApJ, 263, 79
\bibitem[2007]{GAS07}        Gaskell, C.~M., Klimek, E.~S., Nazarova, L.S. arXiv0711.1025
\bibitem[2009]{GAS09} Gaskell, C.~M. 2009, NewAR, 53, 140
\bibitem[2010a]{GAS10a}      Gaskell, C.~M. 2010a, IAUS 267, 203 (arXiv:1003.0036)
\bibitem[2010b]{GAS10b}      Gaskell, C.~M. 2010b, ApJ submitted (arXiv:1008.1057)
\bibitem[2005]{GIL05} Gil-Merino, R., Wambsganss, J., Goicoechea, L.~J., Lewis, G.~F. 2005, A\&A, 432, 83
\bibitem[2009a]{GUT09a}  G{\"u}ltekin,  K., Richstone, D.~O., Gebhardt, K. et al. 2009, ApJ, 698, 198
\bibitem [2004]{HOR04} Horne, K., Peterson, B.~M., Collier, S.~J., Netzer, H.  2004, PASP, 116, 465 
\bibitem[1990]{HIN90}  Hintzen, P., Maran, S.~P., Michalitsianos, A.~G. et al. 1990, AJ, 99, 45
\bibitem[1985]{HUC85}       Huchra, J., Gorenstein, M., Kent, S., et al. 1985, AJ, 90, 691
\bibitem[1994]{HUT94}   Hutsem\'ekers, D., Surdej, J., van Drom, E. 1994, Ap\&SS, 216, 361
\bibitem[2010]{HUT10}  Hutsem\'ekers, D., Borguet, B., Sluse, D., Riaud, P., Anguita, T. 2010, A\&A, 519, 103
\bibitem[2008]{HU08} Hu, C., Wang, J., Ho, L.~C. et al. 2008, ApJ, 683, L115
\bibitem[2006]{JAR06}    Jarvis, M.~J., McLure, R.~J. 2006, MNRAS, 369, 182
\bibitem[1996]{KOL96}  	Kollatschny, W., Dietrich, M. 1996, A\&A, 314, 43
\bibitem[2005]{KAS05} Kaspi, S., Maoz, D., Netzer, H. 2005, ApJ, 629, 61
\bibitem[2007]{KAS07} Kaspi, S., Brandt, W.~N., Maoz, D. et al. 2007, ApJ, 659, 997 
\bibitem[1986]{KAY86} 	Kayser, R., Refsdal, S., Stabell, R. 1986, A\&A, 166, 36
\bibitem[1987]{KEE87}  Keenan, F.~P., Kingston, A.~E., Dufton, P.~L. 1987, MNRAS, 225, 859
\bibitem[2004]{KOC04}     Kochanek, C.~S.  2004, ApJ, 605, 58
\bibitem[1991]{KRO91}  Krolik, J.~H., Horne, K., Kallman, T.~R. et al. 1991, ApJ, 371, 541 
\bibitem[2007]{LAO07} 	Laor, A. 2007, ASPC, 373, 384
\bibitem[1995]{LEW95}  Lewis, G.~F.; Irwin, M.~J. 1995, MNRAS, 276, 103
\bibitem[1996]{LEW96}  Lewis, G.~F.; Irwin, M.~J. 1996, MNRAS, 283, 225	
\bibitem[1998]{LEW98}  Lewis, G.~F.; Belle, K.~E. 1998, MNRAS, 297, 69
\bibitem[2004]{LEW04}     Lewis, G.~F., Ibata, R.~A. 2004, MNRAS, 348, 24
\bibitem[2006]{LEW06}   Lewis, G.~F., Ibata, R.~A. 2006, MNRAS, 367, 1217
\bibitem[1998]{MAG98}     Magain, P., Courbin, F., Sohy, S. 1998, ApJ, 494, 452
\bibitem[2008]{MAR08}   Marconi, A., Axon, D.~J., Maiolino, R. et al. 2008, ApJ, 678, 693
\bibitem[1996]{MAR96} 	Marziani, P., Sulentic, J.~W., Dultzin-Hacyan, D., Calvani, M., Moles, M. 1996, ApJS, 104, 37
\bibitem[2006]{MAR06}  Marziani, P., Dultzin-Hacyan, D., Sulentic, J.~W. 2006 New Developments in Black Hole Research, edited by Paul V. Kreitler, p.123  
\bibitem[2010]{MAR10}	Marziani, P., Sulentic, J.~W., Negrete et al. 2010, MNRAS, 409, 1033
\bibitem[2002]{MCL02} 	McLure, R.~J., Dunlop, J.~S. 2002, MNRAS, 331, 795
\bibitem[2004]{MET04}        Metcalf, R.~B., Moustakas, L.~A., Bunker, A.~J., et al. 2004, ApJ, 607, 43
\bibitem[2010]{MOR10} 	Morgan, C.~W., Kochanek, C.~S., Morgan, N.~D., Falco, E.~E. 2010, ApJ, 712, 1129
\bibitem[2005]{MOR05} Mortonson, M.~J., Schechter, P.~L., Wambsganss, J. 2005, ApJ, 628, 594 
\bibitem[2004]{MOT04} Motta, V., Mediavilla, E., Mu\~noz, J.~A., Falco, E. 2004, ApJ, 613, 86
\bibitem[2009]{MIN09} Minezaki, T., Chiba, M., Kashikawa, N., Inoue, K.~T., Kataza, H. 2009, ApJ, 697, 610
\bibitem[1984]{MUT84}   Mushotzky, R., Ferland, G.~J. 1984, ApJ, 278, 558
\bibitem[1995]{MUR95}   Murray, N., Chiang, J. 1995, ApJ, 454, 105 
\bibitem[1997]{MUR97}   Murray, N., Chiang, J. 1997, ApJ, 474, 91 
\bibitem[2003]{NET03} Netzer, H. 2003, ApJ, 583, L5
\bibitem[2010]{NET10} Netzer, H., Marziani, P. 2010, ApJ, 724, 318
\bibitem[2004]{ONK04} Onken, C.~A., Ferrarese, L., Merritt, D. et al. 2004, ApJ, 615, 645
\bibitem[1991]{PET91} Peterson B.~M., Balonek, T.~J., Barker, E.~S. et al. 1991, ApJ, 368, 119
\bibitem[2005]{PET05} Peterson, B.~M., Bentz, M.~C., Desroches, L.-B. et al. 2005, ApJ, 632, 799 (+Erratum)
\bibitem[2010a]{POI10a}  Poindexter, S., Kochanek, C.~S. 2010a, ApJ, 712, 658 
\bibitem[2010b]{POI10b}  Poindexter, S., Kochanek, C.~S. 2010b, ApJ, 712, 668 
\bibitem[2003]{POP03} 	Popovi\'c, L. {\v C.}, Mediavilla, E.~G., Jovanovi\'c, P., Mu\~noz, J.~A. 2003, A\&A, 398, 975
\bibitem[2004]{POP04}  Popovi\'c, L. {\v C.}, Mediavilla, E., Bon, E., Ili\'c, D. 2004 A\&A, 423, 909
\bibitem[2006]{POP06}	Popovi\'c, L. {\v C.} 2006, SerAJ, 173, 1
\bibitem[2006]{RIC06} 	Rice, M.~S., Martini, P., Greene, J.~E. 2006, ApJ, 636, 654
\bibitem[2002]{RIC02}  	Richards, G.~T., Vanden Berk, D.~E., Reichard, T.~A. et al. 2002, AJ, 124, 1
\bibitem[2004]{RIC04}     Richards, G.~T., Keeton, C.~R., Pindor, B., et al. 2004, ApJ, 610, 679
\bibitem[2010]{RIS10} Risaliti, G., Salvati, M., Marconi, A. 2010, MNRAS, accepted (arXiv1010.2037)
\bibitem[1992]{RIX92} Rix, H.~W., Schneider, D.~P.; Bahcall, J.~N. 1992, AJ, 104, 959
\bibitem[1995]{ROB95} Robinson, A. 1995, MNRAS, 272, 647
\bibitem[1998]{SCH98} Schmidt, R., Webster, R.~L., Lewis, G.~F 1998, MNRAS, 295, 488
\bibitem[1990]{SCH90}   Schneider, P., Wambsganss, J. 1990, A\&A, 237, 42
\bibitem[1973]{SHA73}   Shakura, N.I., Sunyaev, R.A. 1973, A\&A, 24, 337
\bibitem[2007]{SLU07} Sluse, D., Claeskens, J.-F., Hutsem\'ekers, D., et al. 2007, A\&A, 468, 885
\bibitem[2008]{SLU08} Sluse, D., Claeskens, J.-F., Hutsem\'ekers, D., et al. 2008, RMxAC, 32, 83
\bibitem[1999]{SNE99}  Snedden, S.~A., Gaskell, M.C. 1999, ApJ, 521, 91
\bibitem[2005]{SMI05}    Smith, J.~E., Robinson, A., Young, S., Axon, D.~J., Corbett, E.~A. 2005, MNRAS, 359, 846
\bibitem[1995]{SUL95} 	Sulentic, J.~W., Marziani, P., Zwitter, T., Calvani, M. 1995, ApJ, 438, L1
\bibitem[1999]{SUL99}  Sulentic, J.~W., Marziani, P. 1999, ApJ, 518, L9
\bibitem[2000]{SUL00}   Sulentic, J.~W., Marziani, P., Dultzin-Hacyan, D. 2000, ARA\&A, 38, 521
\bibitem[2006]{UDA06}      Udalski, A., Szyma\'nski, M., Kubiak, M., et al. 2006, AcA, 56, 293
\bibitem[2000]{VAN01}  Vanden Berk, D.~E., Richards, G.~T., Bauer, A. et al. 2001, AJ, 122, 549
\bibitem[2000]{VES00}  Vestergaard, M., Wilkes, B.~J., Barthel, P. D 2000, ApJ, 538, L103
\bibitem[2001]{VES01}  Vestergaard, M., Wilkes, B.~J., 2001, ApJS, 134, 1
\bibitem[2006]{VES06}  Vestergaard, M., Peterson, B.~M. 2006, ApJ, 641, 689
\bibitem[1990]{WAM90} Wambsganss, J. 1990 PhD Thesis (Munich University) 
\bibitem[1991]{WAM91} Wambsganss, J., Paczy\'nski, B. 1991, AJ, 102, 864
\bibitem[1992]{WAM92} Wambsganss, J.  1992, ApJ, 386, 19
\bibitem[1994]{WAM94} Wambsganss, J., Paczy\'nski, B. 1994, AJ, 108, 1156
\bibitem[1999]{WAM99} Wambsganss, J. 1999 , JCoAM, 109, 353
\bibitem[2001]{WAM01} Wambsganss, J. 2001, PASA, 18, 207 
\bibitem[2005]{WAY05}	     Wayth, R.~B., O'Dowd, M., Webster, R.~L. 2005, MNRAS, 359, 561
\bibitem[1993]{WIL93} 	Wills, B.~J., Brotherton, M.~S., Fang, D., Steidel, C.~C., Sargent, W.~L.~W. 1993, ApJ, 415, 563
\bibitem[2006]{WIL06} Wilhite, B.~C., Vanden Berk, D.~E., Brunner, R.~J., Brinkmann, J.~V.  2006, ApJ, 641, 78
\bibitem[1994]{WIT94} 	Witt, H.~J., Mao, S. 1994, ApJ, 429, 66
\bibitem[1985]{WHI85} Whittle, M. 1985, MNRAS, 213, 33
\bibitem[2007]{YOU07} Young, S. Axon, D.~J., Robinson, A., Hough, J.~H., Smith, J.~E. 2007, Nature, 450, 74
\bibitem[1991]{YEE91}   Yee, H.~K.~C., De Robertis, M.~M. 1991, ApJ, 381, 386
\bibitem[2008]{ZAM08}  	Zamfir, S., Sulentic, J.~W., Marziani, P. 2008, MNRAS, 387, 856
\bibitem[2009]{ZHU09}  	Zhu, L., Zhang, S.~N., Tang, S. 2009, ApJ, 700, 1173

\end{thebibliography}
\end{document}